\documentclass[namedreferences]{solarphysics}
\usepackage[hyperref,optionalrh,solaromanenum]{spr-sola-addons} 
\usepackage{epsfig}                     
\usepackage{grffile}
\usepackage{graphicx,subfig}        
\usepackage{courier}         
\usepackage{amssymb}        
\usepackage{color}           
\usepackage{breakurl}        

\usepackage{courier}         
\usepackage{amssymb}         
\usepackage{color}           
\usepackage{breakurl}        

\chardef\us=`\_

\begin{document}

\begin{article}

\begin{opening}

\title{Visibility and Origin of Compact Interplanetary Radio Type~IV Bursts} 
\author[addressref={aff1},corref,email={natash@utu.fi}]{\inits{N.}\fnm{Nasrin}~\lnm{Talebpour Sheshvan}}
\author[addressref={aff2,aff1},email={silpoh@utu.fi}]{\inits{S.}\fnm{Silja}~\lnm{Pohjolainen}}

\address[id=aff1]{Department of Physics and Astronomy, University of Turku, Turku, Finland}
\address[id=aff2]{Tuorla Observatory, Department of Physics and Astronomy, University of Turku, Turku, Finland}
\runningauthor{Talebpour Sheshvan and Pohjolainen}
\runningtitle{Visibility of Compact IP type IV bursts}


\begin{abstract}
We have analysed radio type IV bursts in the interplanetary (IP) space
at decameter--hectometer (DH) wavelengths, to find out their source
origin and a reason for the observed directivity. We used radio
dynamic spectra from the instruments on three different spacecraft,
STEREO-A, {\it Wind}, and STEREO-B, that were located approximately 90
degrees apart from each other in 2011-2012, and thus gave a 360 degree
view to the Sun. The radio data was compared to white-light and
extreme ultraviolet (EUV) observations of flares, EUV waves, and
coronal mass ejections (CMEs) in five solar events. We find that the
reason for observing compact and intense DH type IV burst emission
from only one spacecraft at a time is due to the absorption of
emission to one direction and that the emission is blocked by the
solar disk and dense corona to the other direction. The geometry also
makes it possible to observe metric type IV bursts in the low corona
from a direction where the higher-located DH type IV emission is not
detectable. In the absorbed direction we found streamers present, and
these were estimated to be the locations of type II bursts, caused by
shocks at the CME flanks. The high-density plasma was therefore most
probably formed by shock--streamer interaction. In some cases the type
II-emitting region was also capable of stopping later-accelerated
electron beams, visible as type III bursts that ended near the type II
burst lanes.
\end{abstract}
\end{opening}


\section{Introduction}

Coronal mass ejections (CMEs) are a large scale phenomena of plasma
and magnetic field explosion from the Sun into the interplanetary (IP)
medium. Solar events such as flares and CMEs accelerate particles,
with different mechanisms, and cause them to propagate from the solar
corona into the IP space. For reviews see, {\it e.g.}, \cite{pick08}
and \cite{nindos08}. The particle paths, their directions and
locations, can be studied using radio bursts at a wide wavelength
range, from metric to kilometric wavelengths.

Solar radio bursts in the plasma regime (below $\approx$ 1 GHz) are
generally classified into different types, based on their appearance
in the dynamic spectrum and their emission mechanism. The most common
types are radio type II and type III bursts. Type II bursts are
thought to be due to propagating shock fronts that accelerate
electrons, and type III bursts are caused by propagating electron
beams.  Both types of emission are produced when supra-thermal 
electrons interact with the surrounding plasma. This creates radio
emission at the local plasma frequency $f_{p}$ and/or its
harmonics. The particles in type III bursts are typically
flare-accelerated electrons travelling along coronal magnetic field
lines \citep{krupar2018} while type II bursts have been
associated with CME-induced shock waves \citep{cane84,leblanc01} .

The more rare radio type IV bursts have been closely associated with
expanding and/or rising magnetized plasma structures. The
emission mechanism can therefore be both synchrotron emission by
trapped high-energy electrons gyrating in the magnetic field and
plasma emission as the magnetic cloud lifts off from the Sun.
Typically metric type IV bursts are divided into two categories,
stationary and moving. The characteristics of stationary type IV
bursts have been defined by \cite{kundu65}, to be located near a
flare, to be strongly circularly polarized, smooth and broad-band, and
located low in the corona near the corresponding plasma level. The
stationary type IV bursts show no systematic movement and the bursts
may also occur without any associated type II emission. It has been
suggested that the source of emission is particles accelerated in a
flare, trapped in post-flare loops and arcades. Moving type IV bursts
on the other hand show definite frequency drifts toward the lower
frequencies, and they are mostly thought to represent particles
trapped in rising CME structures. \cite{kundu65} also stated that type
IV bursts show directivity in their emission. Some type IV bursts
extend in the dynamic spectra to decameter--hectometer (DH)
wavelengths and they can be recorded by space-borne instruments at
frequencies below $\approx$ 20 MHz (Earth ionosphere limit). The
characteristics of IP moving type IV bursts have recently been listed
and studied by \cite{hillaris2016}, and their catalogue contains
bursts recorded in 1998--2012 by the radio experiment on-board {\it
  Wind} satellite. Directivity in IP type IV burst emission was later
noticed by \cite{gopal2016}.

Closely associated with high-speed CMEs are EUV-waves, observed in the
extreme ultra-violet wavelength range in the low corona. These bright
features and the dimming behind them typically start from the flare
location and propagate globally over the solar disk. However, no
strong correlation was found between the global EUV disturbance speeds
and flare intensities, or CME magnitudes \citep{nitta2013}.  Recently,
\cite{kwon2017} found that halo-CME widths are in agreement with the
widths of EUV waves in the low corona, suggesting a common origin for
these structures. In some cases the heights of EUV waves have been
found to be compatible with the heights of radio type II burst sources
\citep{cunha2015}. Since radio type IV bursts can be observed over a
wide frequency range, the result of the studies made over the past
decades has been to define type IV bursts as any broad-band,
long-duration, continuum-like emission occurring after a
flare. Moreover, imaging has only been available for
decimetric--metric bursts that occur in the low corona, but not for DH
type IV bursts in the IP space.

In 2007 two more spacecrafts with radio spectrograph were launched to
orbit the Sun, and with {\it Wind}, STEREO A and STEREO B instruments
a 3D-view of the Sun was made possible, if not real imaging.  In this
study we use the radio data from these three spacecraft and the
questions we try to answer are, where are IP type IV bursts located,
is there directivity in their emission, and are they a continuation of
metric type IV burst emission.

\section{Data Analysis}

\begin{figure}[!h]
   \centering
   \includegraphics[width=0.48\textwidth]{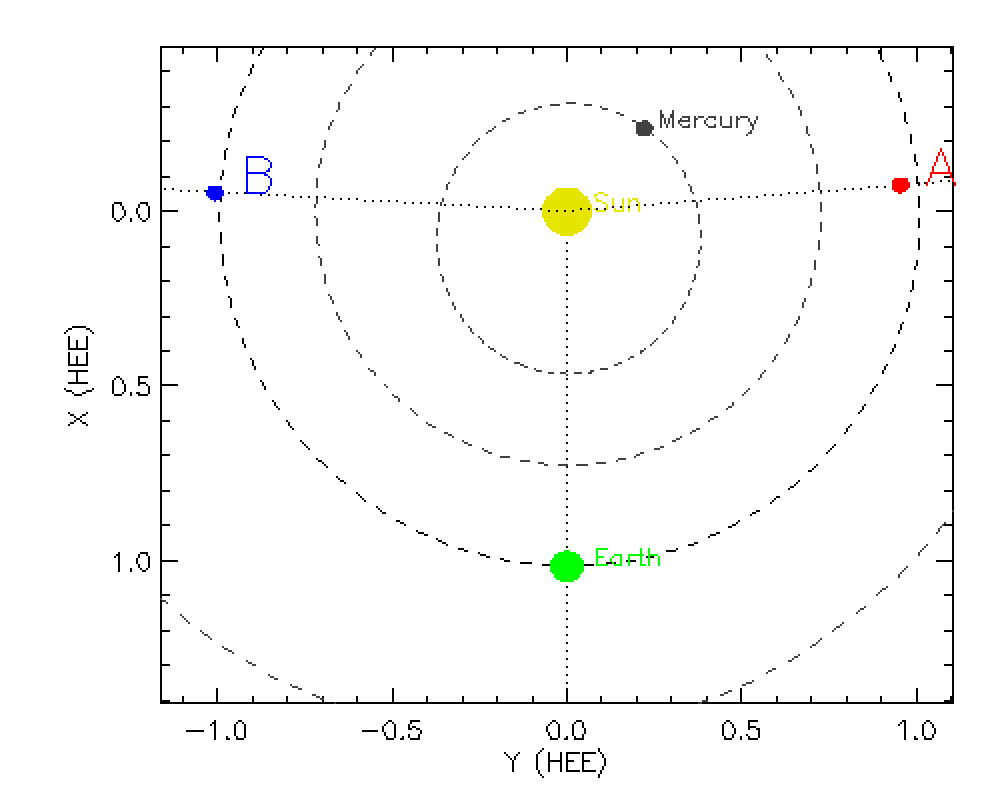}
   \includegraphics[width=0.48\textwidth]{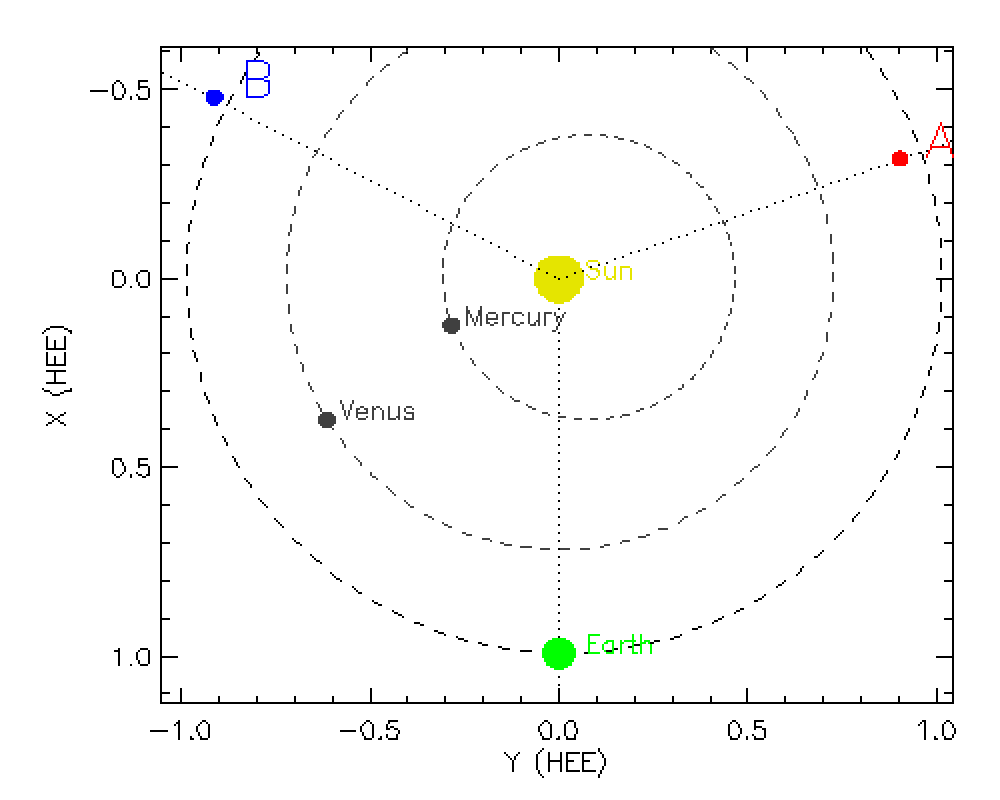}
   \caption{Locations of STEREO A (red circle) and B (blue circle) spacecraft
relative to the Earth (green circle) on 4 June 2011 (left) and 5 March 2012 (right). 
These are the first and last event dates in our sample, and the plots show
how the orbital distance was increasing in respect to Earth and the {\it Wind}
satellite located near L1. During this time period the Sun could be viewed
in 3D. (\url{https://stereo-ssc.nascom.nasa.gov})
} 
\label{stereoview}%
\end{figure}

The time period to find DH type IV bursts was set to 2011--2012 in
order to have a full 3D-view of the Sun, as then the spacecraft were
orbiting the Sun with an $\approx$\,90$^{\circ}$ angular separation,
and the 'backside' of the Sun could be imaged in EUV and white-light
with the instruments on-board STEREO A and B, see Figure~\ref{stereoview}.
The Earth-side imaging were provided by {\it Solar Dynamic Observatory}
(SDO) and {\it Solar and Heliospheric Observatory} (SOHO).

The DH type IV bursts were selected from the catalogue of solar type
II and type IV bursts detected by the WAVES instruments on {\it Wind}
\citep{bougeret95} and STEREO A and B \citep{bougeret2008}. The catalogue
has been prepared by Michael L. Kaiser. Altogether eleven DH type IV events
were found from this time period but only five of them were found to show a
compact, intense, and long-duration DH type IV burst, with radio and
imaging observations from all the three viewing angles. These five
events and their source origins are listed in Tables~\ref{events} and
\ref{flares}. Figure~\ref{stereoview} shows the spacecraft positions
relative to the Sun, on the first and last event date.

The highest observing frequencies for {\it Wind}/WAVES and STEREO/WAVES
are 14 MHz and 16 MHz, respectively. In the case of plasma emission these
frequencies correspond to densities of 2.4\,$\times$\,10$^6$ cm$^{-3}$
and 3.2\,$\times$\,10$^6$ cm$^{-3}$ (plasma emission frequency $f_p$
$\approx$ 9000\,$\sqrt{n_e}$, where $f_p$ is in Hz and ${n_e}$ in
cm$^{-3}$).

Three of the DH type IV bursts were preceded by metric type IV bursts
that were observed with ground-based instruments from Earth. The other
two DH type IVs had their source origin on the backside of the Sun and
therefore the radio emission at lower atmospheric heights could not
be observed (Table~\ref{flares}).
Two of the five flares originated from the backside of the Sun, with
no X-ray observations, and therefore their GOES X-ray flare class is
not known. The other three were GOES X-class flares.


\begin{table}[!hb]
\caption{DH type IV bursts observed from three different viewing angles.}
\label{events}              
\begin{tabular}{l l l l l}
\hline       
Date       & Type IV        &                    & Type IV       & \\
yyyy-mm-dd & appearance$^1$ & STEREO-B           & Wind          & STEREO-A \\
           & UT             & (source loc.)      & (source loc.) & (source loc.)\\
\hline      
2011-06-04 & 07:12--09:00  & --                   & --           & strong (W50)\\
2011-06-04 & 22:15--23:30$^2$ & --                & --           & strong (W70)\\
2011-09-22 & 11:10--12:40  & strong (W10)         & --           & -- \\
2012-01-27 & 18:30--20:10  & -- \, \, \, (N-limb) & --           & strong (E50)\\
2012-03-05 & 04:15--06:00  & faint  \, (W80)      & strong (E50) & -- \\ 
\hline                  
\end{tabular}\\
$^1$Time of IP type IV start and end at 16--14 MHz (instrument limit).\\
$^2$First part, a second more narrow-band enhancement follows at 23:45--01:20 UT.
\end{table}


\begin{table}[!hb]
\caption{Flares and metric type IV bursts prior to the DH type IV bursts.}
\label{flares}              
\begin{tabular}{l l l l l l}
\hline       
Date       & Flare  & Flare    & GOES  & Flare        & Metric  \\
yyyy-mm-dd & start  & maximum  & class & location     & type IV \\
           & UT     & UT       &       & (Earth view) &      \\
\hline      
2011-06-04 & 06:20 &       &      & N20W140           & not available\\
2011-06-04 & 21:45 &       &      & N20W160           & not available\\
2011-09-22 & 10:29 & 11:01 & X1.4 & N13E78            & 10:40--13:40\\
2012-01-27 & 17:37 & 18:37 & X1.7 & N27W71            & 18:15--18:35 \\
2012-03-05 & 02:30 & 04:09 & X1.1 & N17E52            & 04:20--05:20 \\ 
\hline                  
\end{tabular}\\
\end{table}


All the type IV events were associated with EUV waves and they are listed
in Table~\ref{waves}. All the observed strong, long-duration type IV
bursts were associated with global EUV waves that were observed to
cross the solar disk in this viewing angle. The EUV waves that were
visible only near the limb, or which covered only a small portion of
the solar disk, had either no corresponding type IV emission or the
type IV was faint in intensity and ended earlier.

The associated CMEs and type II shocks are listed in Table~\ref{cmes-and-2s}.
All the CMEs had very high speed, in the range of 1600--2900 km s$^{-1}$ near
the time of type IV burst appearance (2nd order fit to the observed plane-of-sky
heights, from the LASCO CME Catalogue). The CME on 27 January 2012 was the only
one accelerating while all the other CMEs were decelerating. 


\begin{table}[!h]
\caption{EUV waves associated with the DH type IV events}
\label{waves}              
\begin{tabular}{l l r r r}
\hline       
Date       & EUV         &                       & EUV wave (radio)         & \\
yyyy-mm-dd & wave$^1$    & STEREO-B/EUVI         & SDO/AIA                  & STEREO-A/EUVI \\
           & km s$^{-1}$  & (WAVES)               & ({\it Wind}/WAVES)       & (WAVES) \\
\hline      
2011-06-04 &            & -- \, \, \, \, (no IV) & -- \, \, \, \, (no IV)   & global (strong IV) \\
2011-06-04 &            & at limb \, \, (no IV)  & -- \, \, \, \, (no IV)   & global (strong IV) \\
2011-09-22 & 595        & global (strong IV)     & at limb \, \, (no IV)    & -- \, \, \, \,  (no IV) \\
2012-01-27 & 635        & N-quarter (no IV)      & W-quarter (no IV)        & global (strong IV) \\
2012-03-05 & 915        & half-disk (faint IV)   & global (strong IV)       & -- \, \, \, \, (no IV)\\
\hline                  
\end{tabular}\\
$^1$Large-scale Coronal Propagating Fronts observed by SDO/AIA \citep{nitta2013}, note
that the velocity value comes from Earth-view observations only.
\end{table}

\begin{table}[!h]
\caption{CMEs and type II bursts associated with the DH type IV events.}
\label{cmes-and-2s}              
\begin{tabular}{l c c c c c c c}
\hline       
Date       & CME        & CME      & Type IV & Type IV  & Type II    & Type II   & CME  \\
yyyy-mm-dd & height$^1$ & speed$^2$ & lowest  & lowest  & freq.$^3$  & height$^4$& height$^5$\\
           & R$_{\odot}$ & km s$^{-1}$ & UT     & MHz     & MHz       & R$_{\odot}$ & R$_{\odot}$ \\
\hline      
2011-06-04 & 6.6       & 1700      & 08:00    & 6      & 1.6       &  8.0    & 12.8 \\
2011-06-04 & 6.3       & 2900      & 23:05    & 6      & 0.8       & 11.7    & 17.8 \\
2011-09-22 & 8.4       & 2200      & 12:15    & 8      & 0.7       & 12.6    & 18.5 \\
2012-01-27 & 3.8       & 2000      & 19:30    & 9      & 1.4       &  8.3    & 16.1 \\
2012-03-05 & 6.6       & 1600      & 05:15    & 7      & 1.0       & 10.4    & 14.8 \\
\hline                  
\end{tabular}\\
$^1$CME height at the time of the type IV appearance, interpolated or extrapolated 
when not observed at that time. The type IV appearance at 16--14 MHz (instrument limit)
corresponds to a height of 2--3~R$_{\odot}$, depending on the atmospheric density model.\\
$^2$CME speed from a 2nd order fit to the observed plan-of-sky heights near the time of
type IV burst appearance. \\
$^3$Type II burst frequency at the time of lowest type IV frequency.\\ 
$^4$Type II burst height from the density model of \cite{vrsnak04}.\\ 
$^5$CME height at the time of the lowest type IV frequency, interpolated or extrapolated 
when not observed at that time. 
\end{table}

The lowest type IV frequencies mark the time when the particle supply ends
and/or the trapping loops start to shrink. The lowest observed type IV
frequencies were 9--6 MHz, which correspond to heliocentric heights of
3.4--4.1 R$_{\odot}$ when calculated using the atmospheric density model of
\cite{vrsnak04}. The type II bursts were at that time already at much lower
frequencies, at 1.6--0.7 MHz, which correspond to heights of 8.0--12.6 R$_{\odot}$.
The height separation between the front of the type IV emission source and
the propagating type II shock was therefore in the range of 4--9 R$_{\odot}$.

When comparing the type II heights with the simultaneous CME leading
front heights, it is evident that the type II bursts were not due to
CME bow shocks, but were shocks at the CME flanks (height separation
of about 4--8 R$_{\odot}$).

\begin{figure}[!ht]
   \centering
 \includegraphics[width=0.45\textwidth]{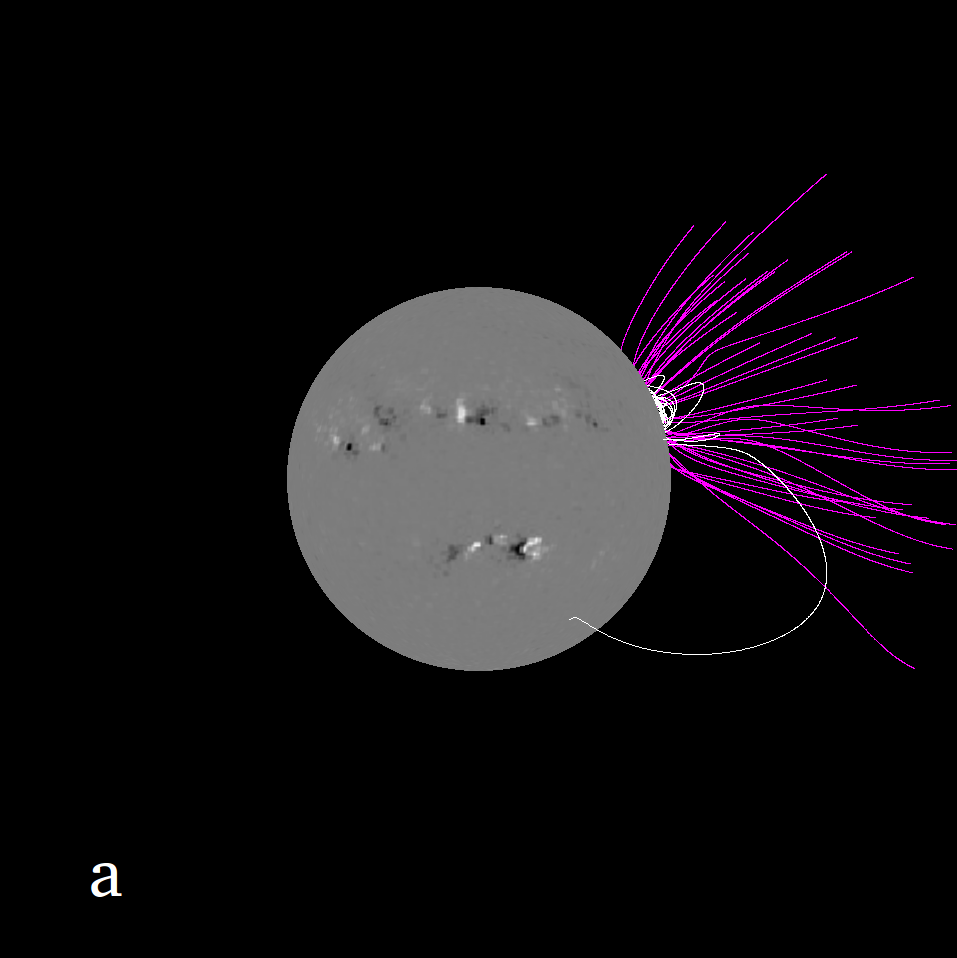}
  \includegraphics[width=0.45\textwidth]{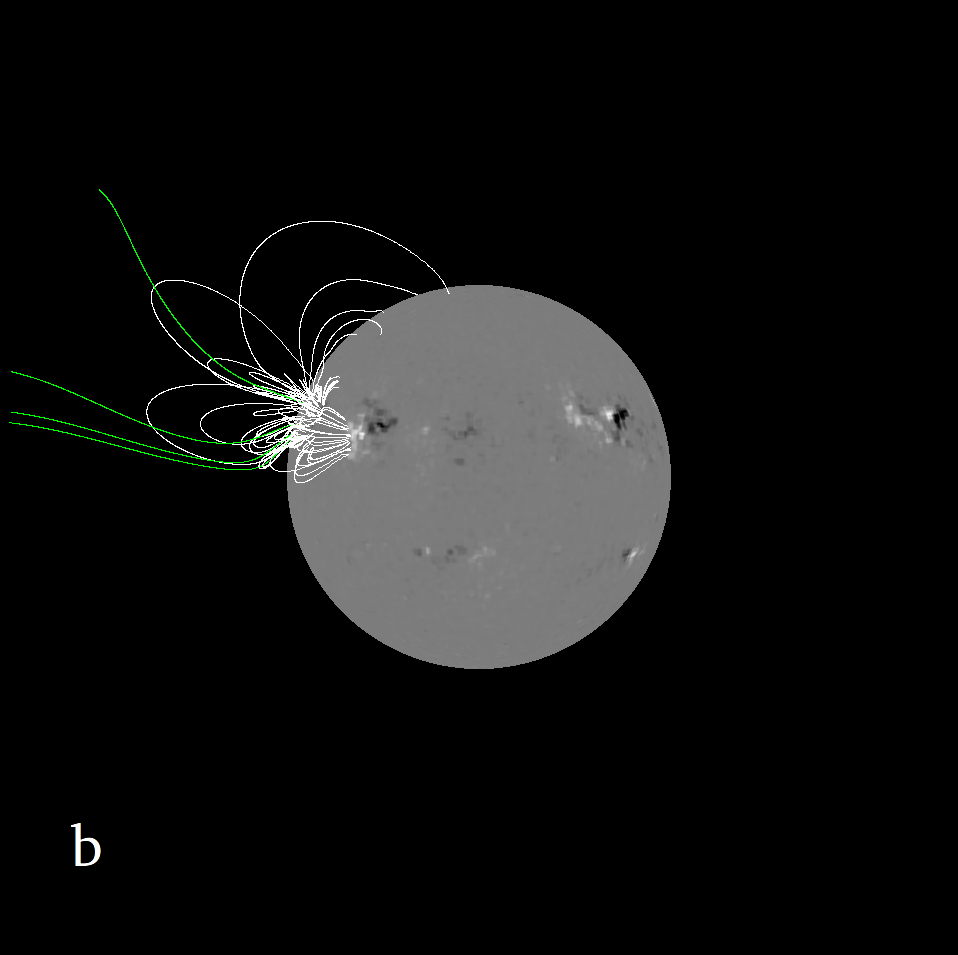}
  \includegraphics[width=0.45\textwidth]{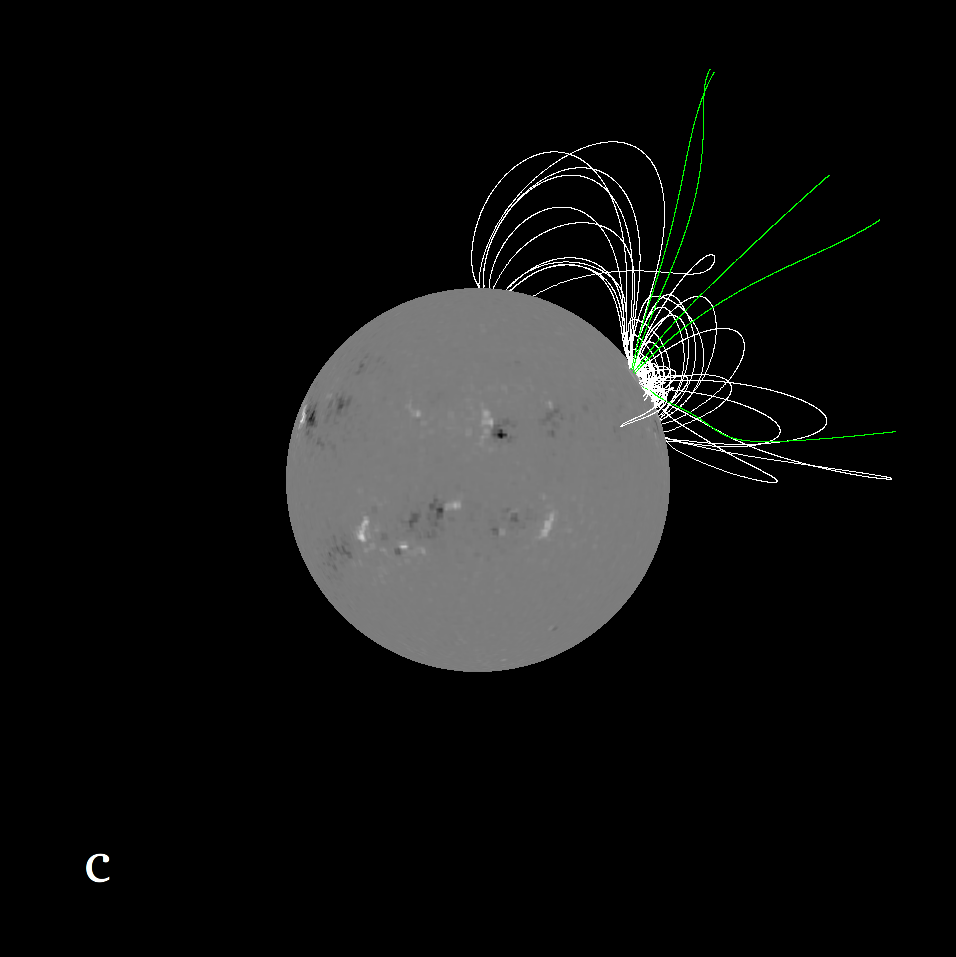}
 \includegraphics[width=0.45\textwidth]{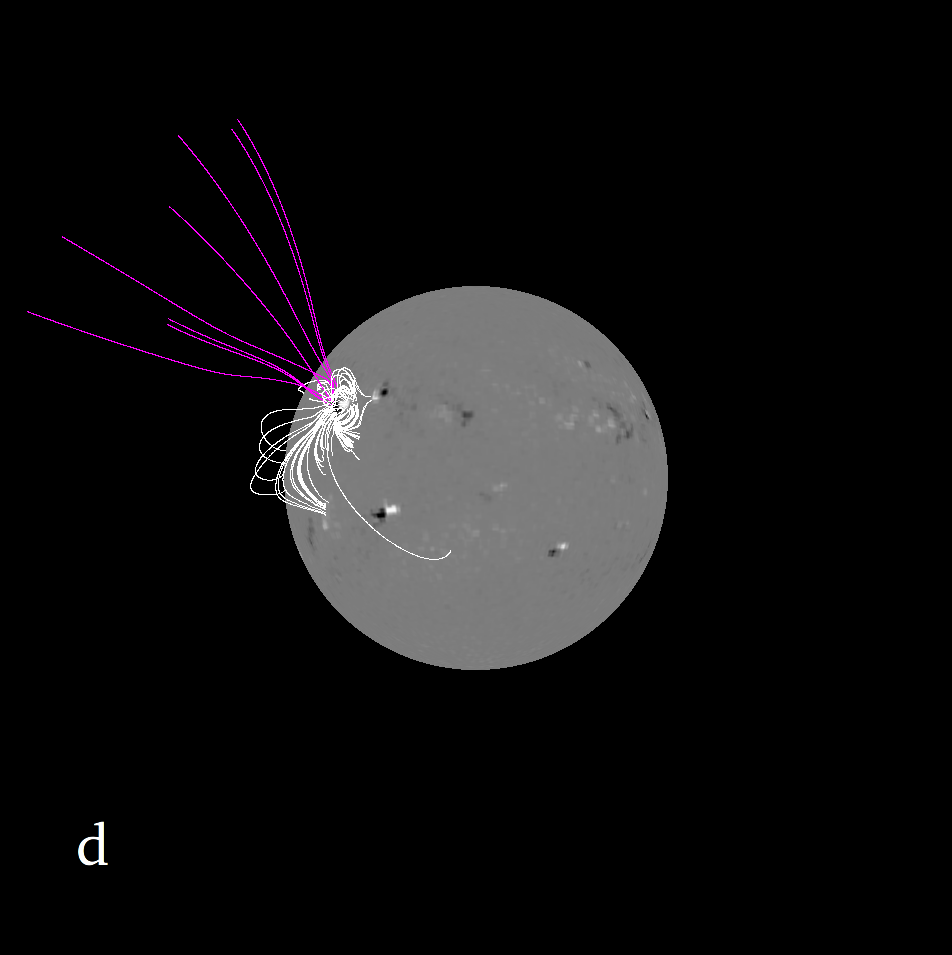}
 \caption{The Potential Field Source Surface (PFSS) maps made from the 
   SOHO/MDI magnetograms (Earth view from L1) provide an approximation of
   the magnetic field line structures in each event, on a) 4 June 2011,
   b) 22 September 2011, c) 27 January 2012, and d) 5 March 2012.  
   In the above maps, the purple and green lines are open field lines indicating
   negative and positive polarity, respectively. The white lines are closed
   field lines. Open field lines show locations where accelerated particles
   can escape from the active region, for example electron beams observed
   as type III bursts.  }
\label{pfss}%
\end{figure}
\begin{figure}[!ht]
   \centering
  \includegraphics[width=0.45\textwidth]{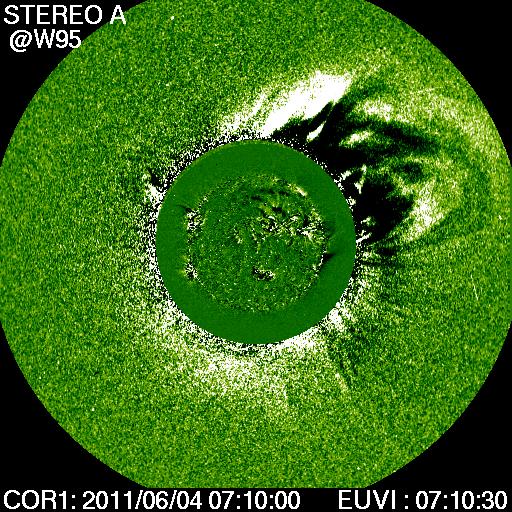} 
   \includegraphics[width=0.45\textwidth]{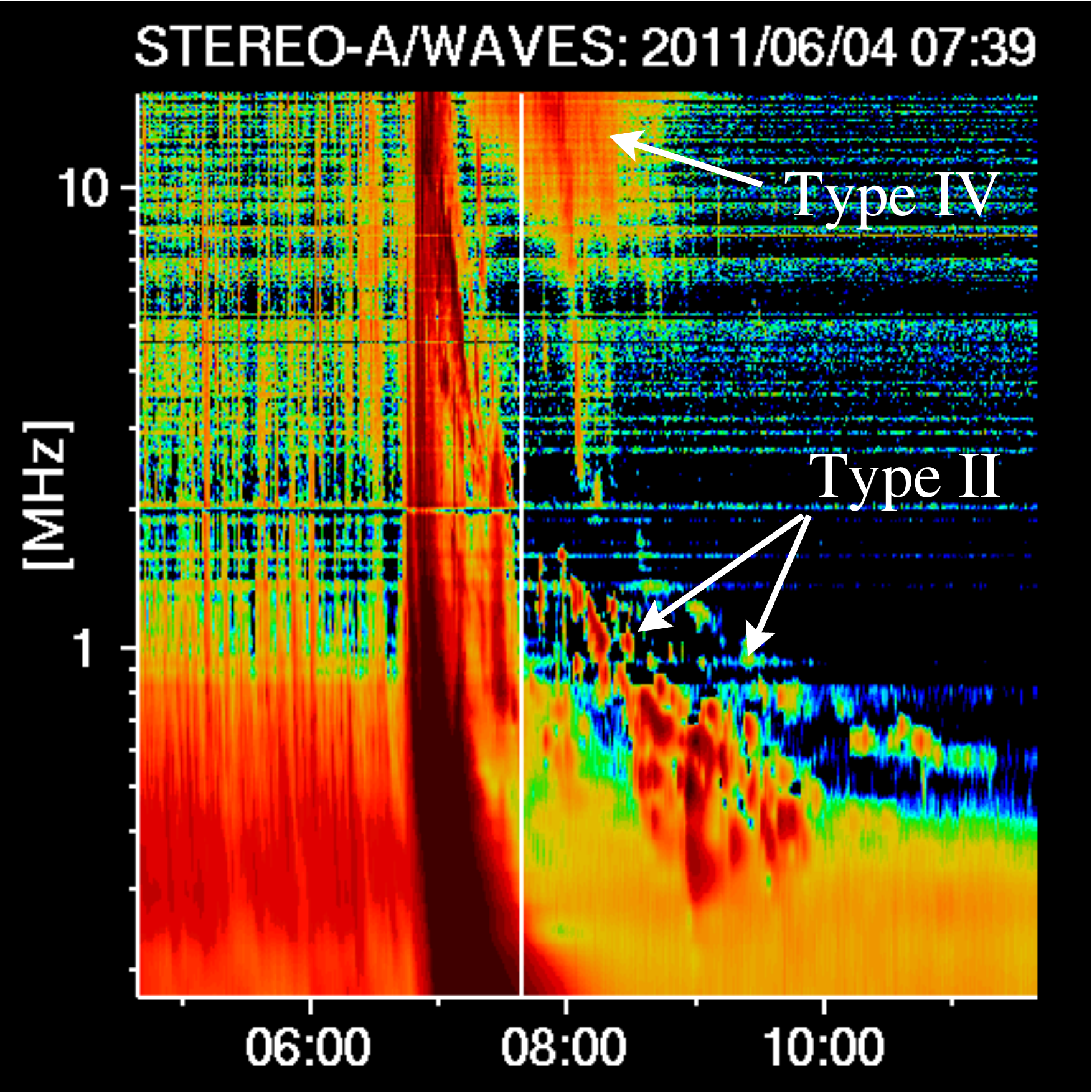}
 \includegraphics[width=0.45\textwidth]{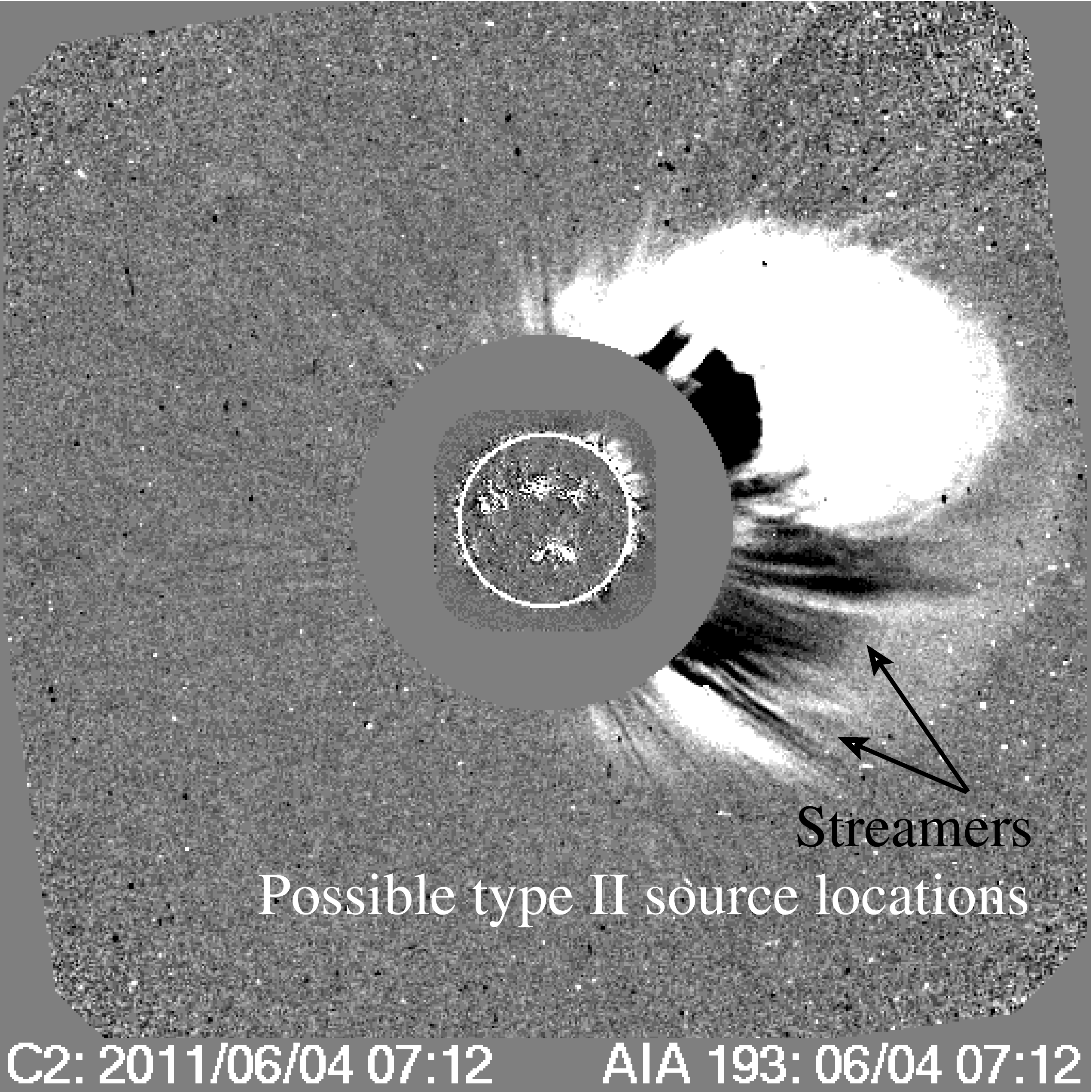}
 \includegraphics[width=0.45\textwidth]{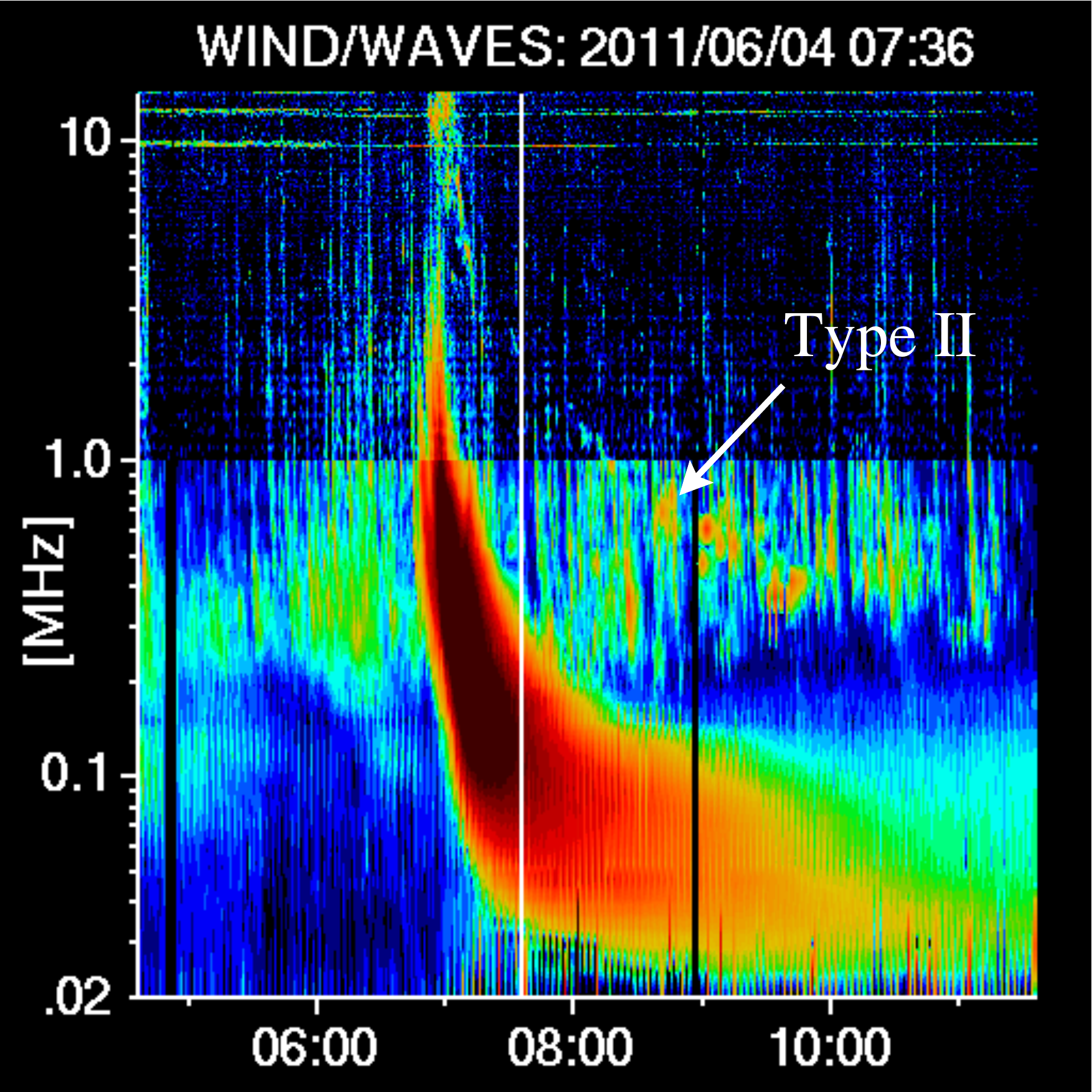}
  \includegraphics[width=0.45\textwidth]{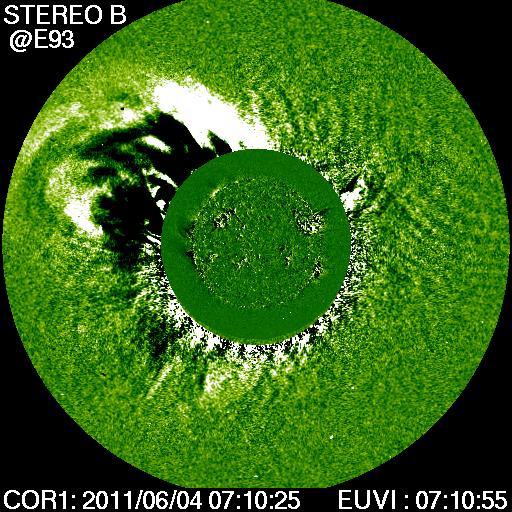}
  \includegraphics[width=0.45\textwidth]{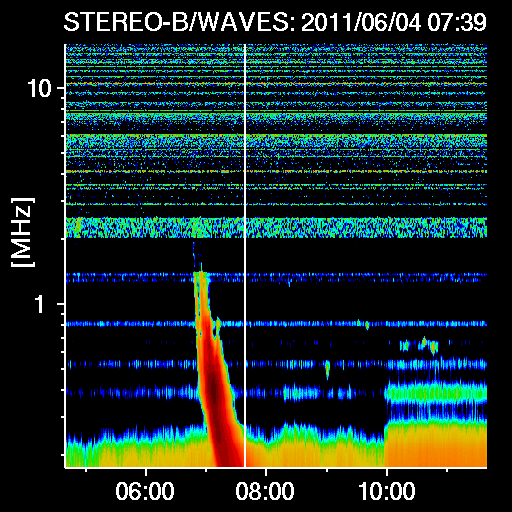}
   \caption{Solar event on 4 June 2011: Flare location on the backside of the Sun, near N20W140
(Earth view). 
} 
\label{2011june4A}%
\end{figure}
\begin{figure}[!ht]
   \centering
  \includegraphics[width=0.45\textwidth]{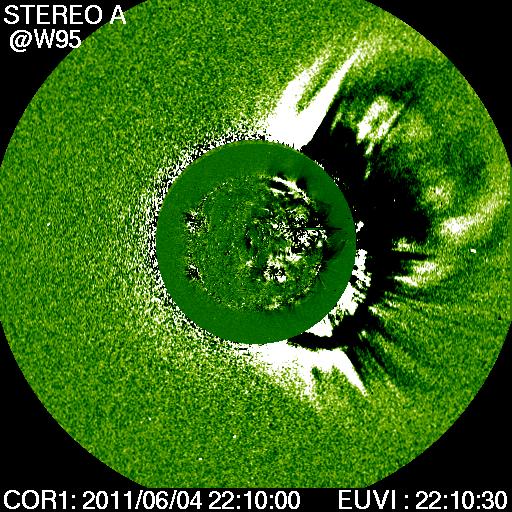}  
   \includegraphics[width=0.45\textwidth]{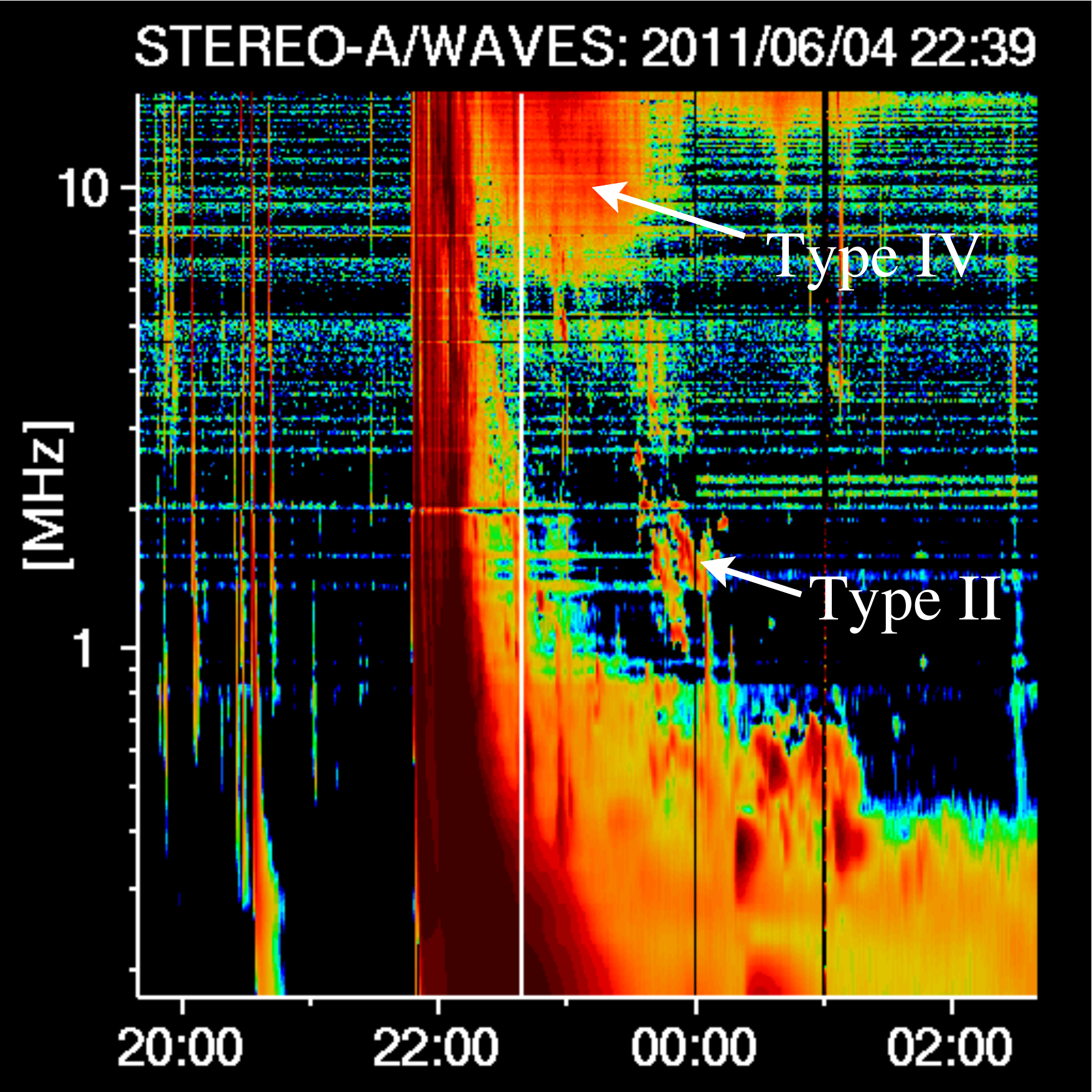}
 \includegraphics[width=0.45\textwidth]{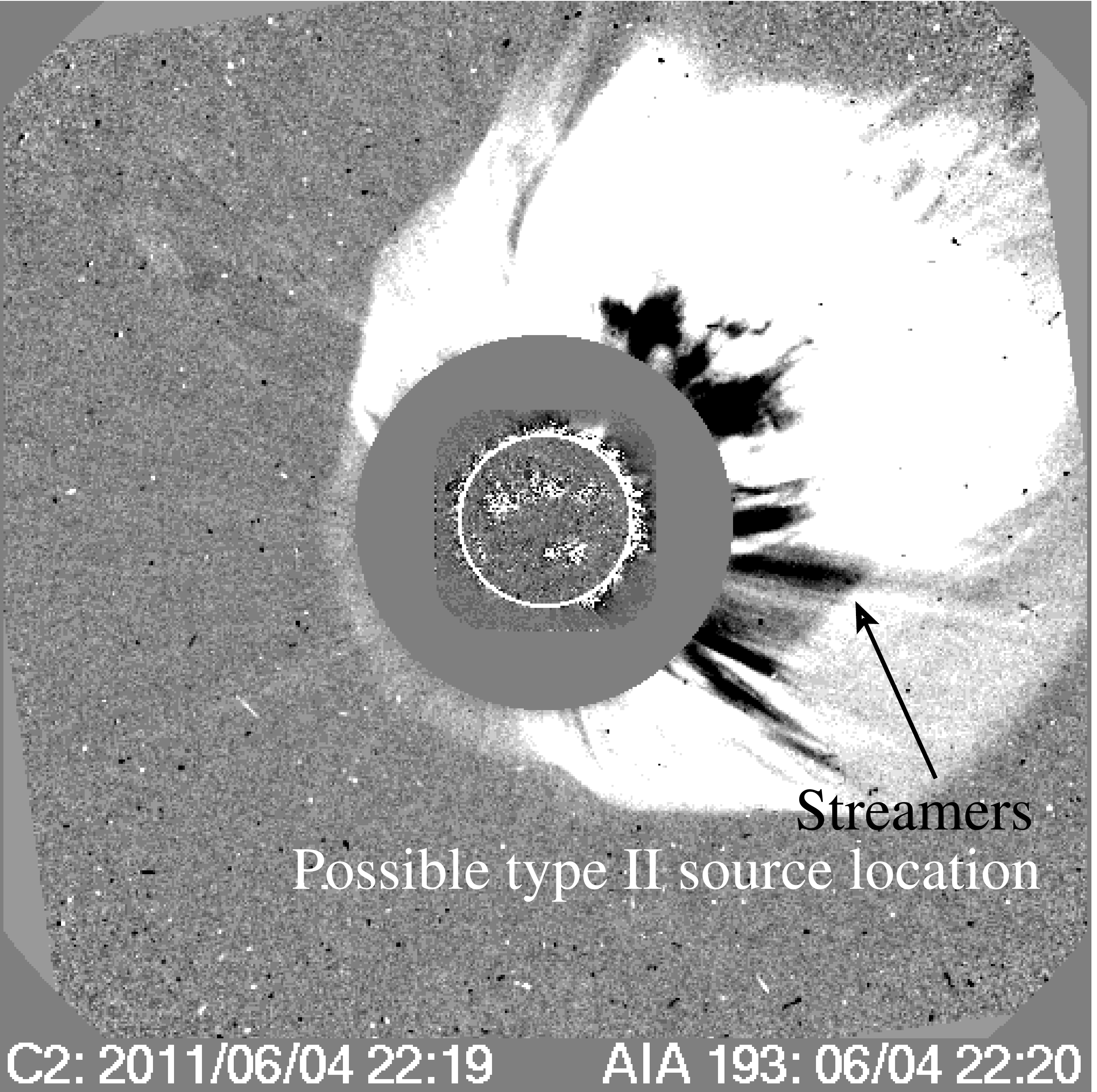}
 \includegraphics[width=0.45\textwidth]{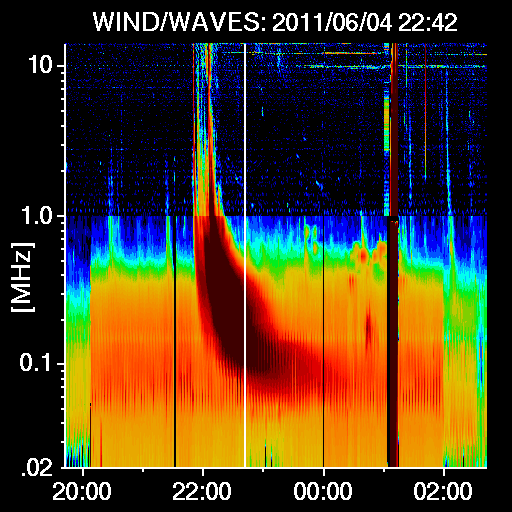}
 \includegraphics[width=0.45\textwidth]{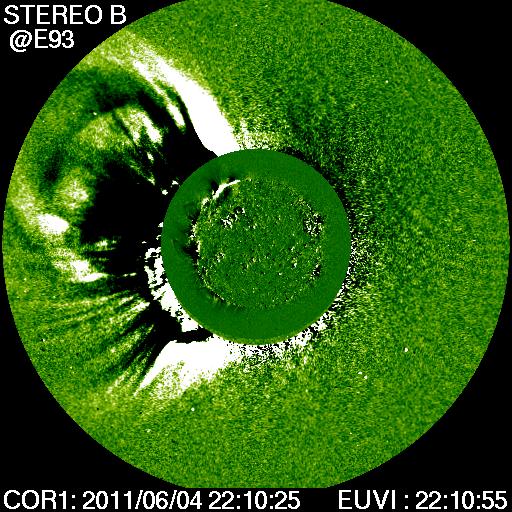}
  \includegraphics[width=0.45\textwidth]{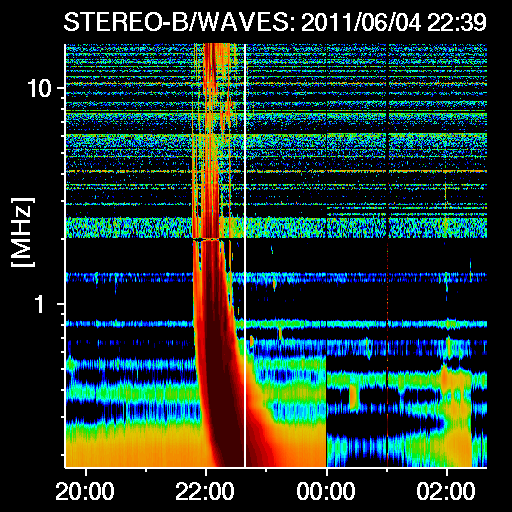}
  \caption{Solar event on 4 June 2011: Flare location on the backside of the Sun,
    near N20W160 (Earth view). 
} 
\label{2011june4B}%
\end{figure}
\begin{figure}[!h]
   \centering
 \includegraphics[width=0.45\textwidth]{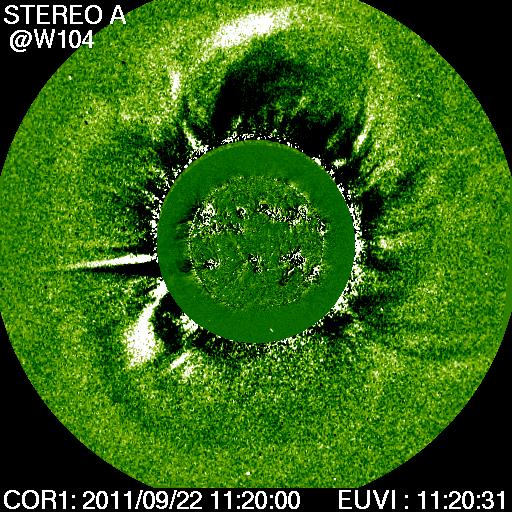}
  \includegraphics[width=0.45\textwidth]{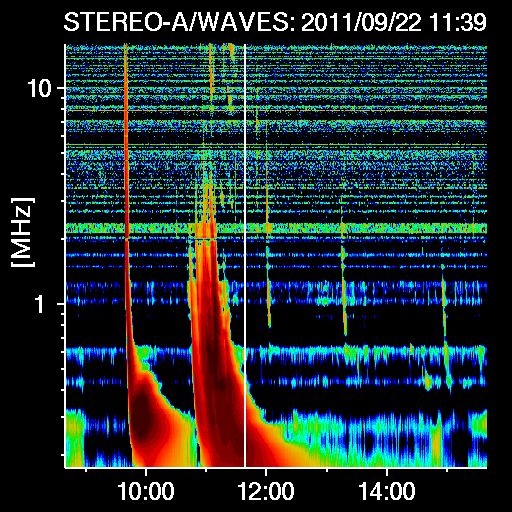} 
  \includegraphics[width=0.45\textwidth]{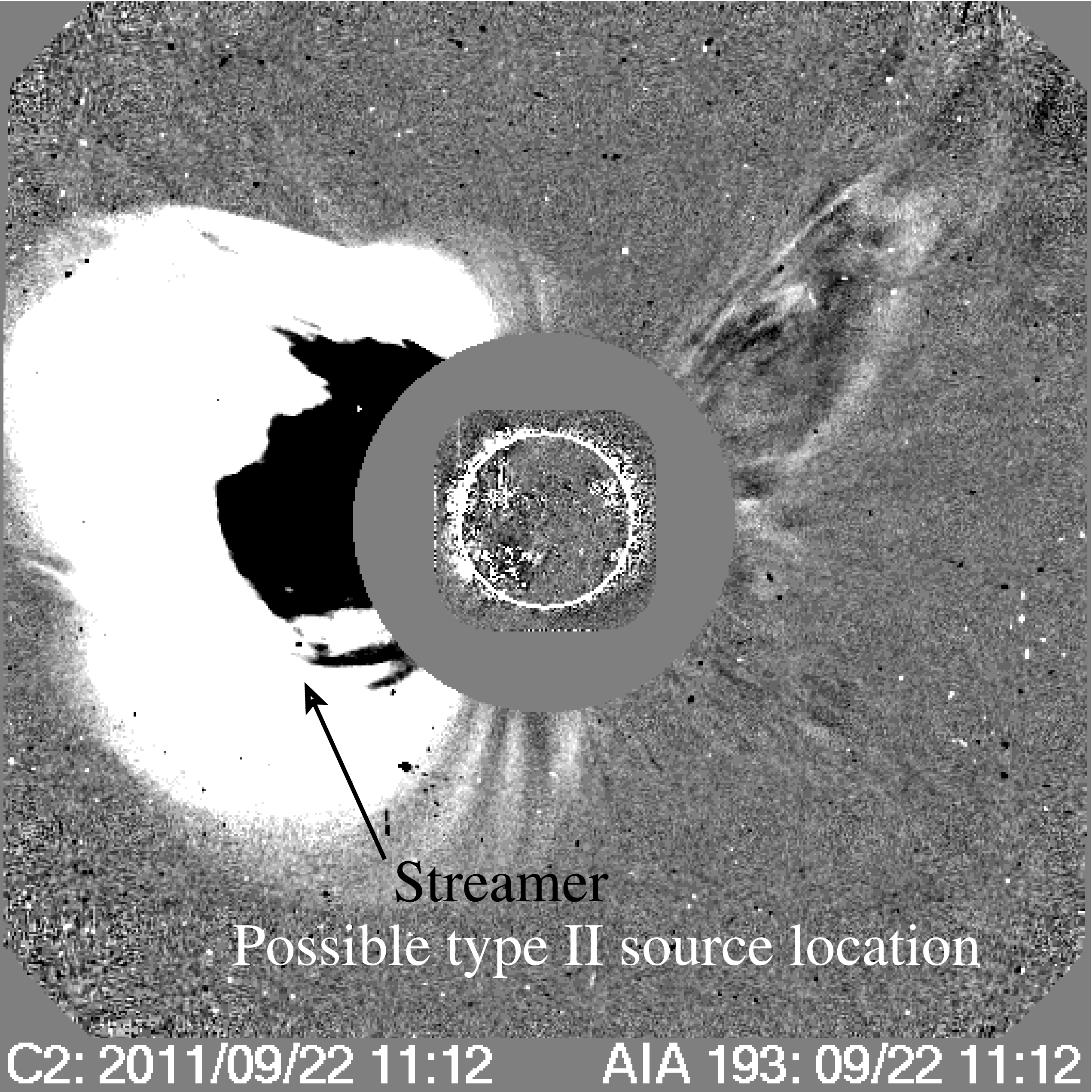}
  \includegraphics[width=0.45\textwidth]{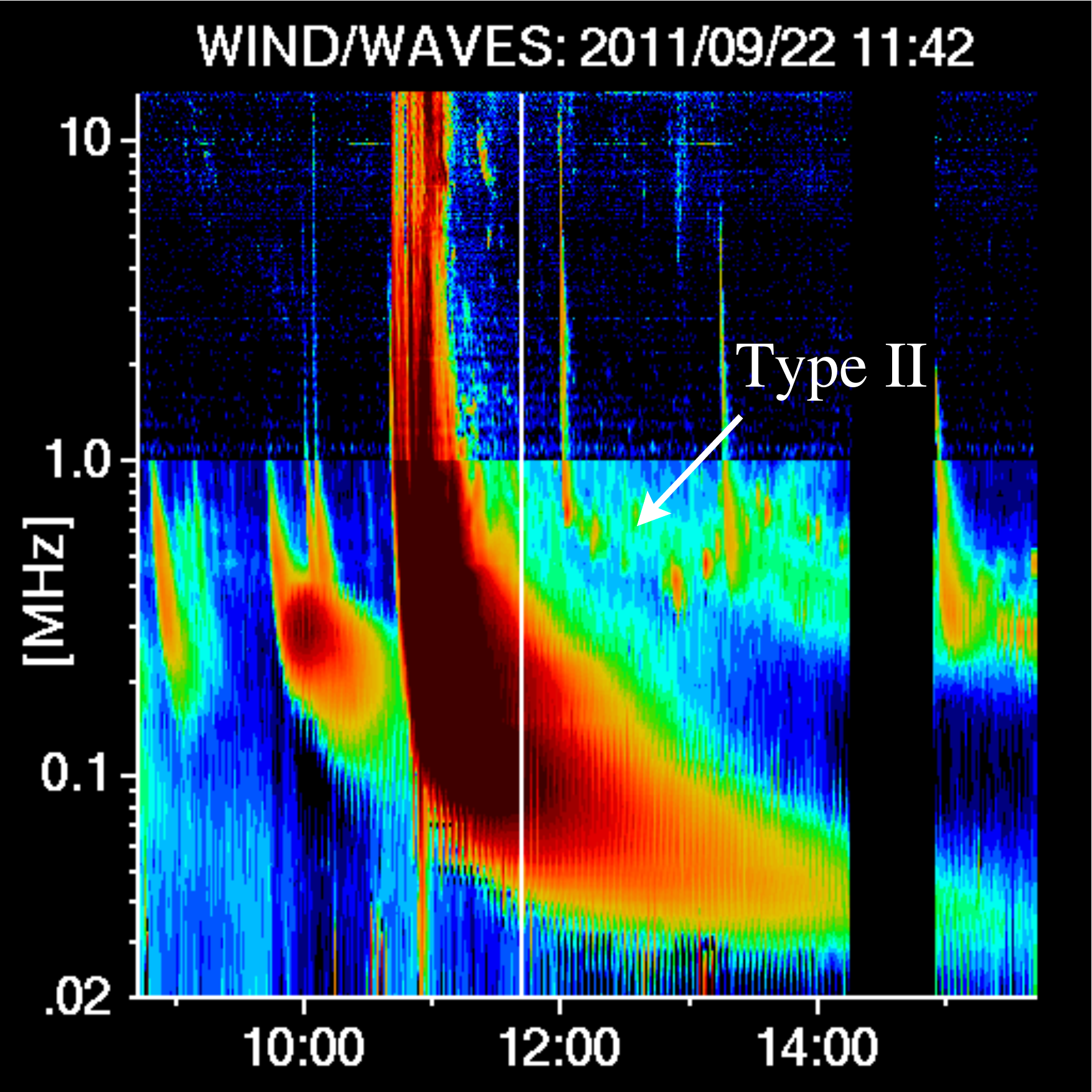}
  \includegraphics[width=0.45\textwidth]{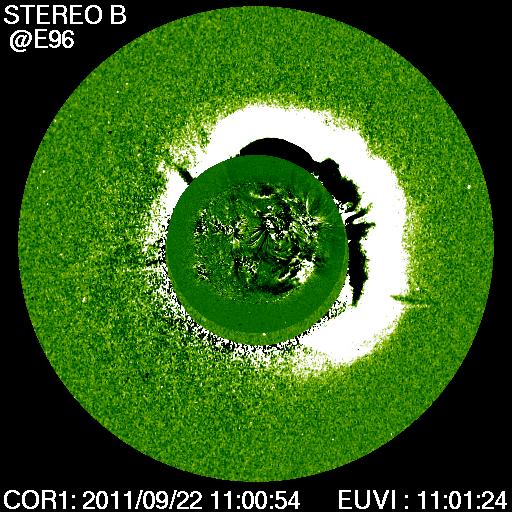}  
   \includegraphics[width=0.45\textwidth]{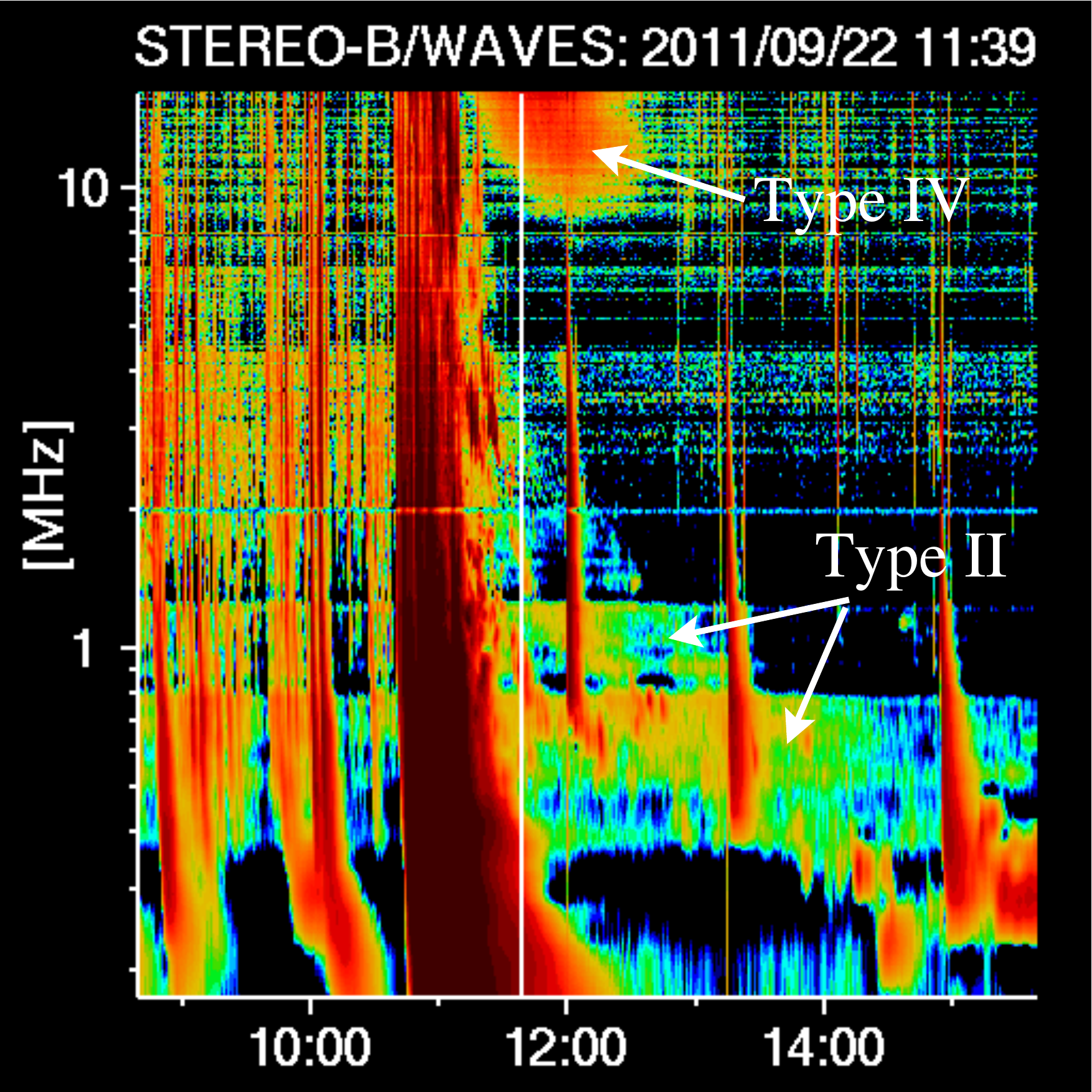} 
  \caption{Solar event on 22 September 2011: Flare location at N13E78 in AR1302.
} 
\label{2011sep22}%
\end{figure}
\begin{figure}[!ht]
   \centering
   \includegraphics[width=0.45\textwidth]{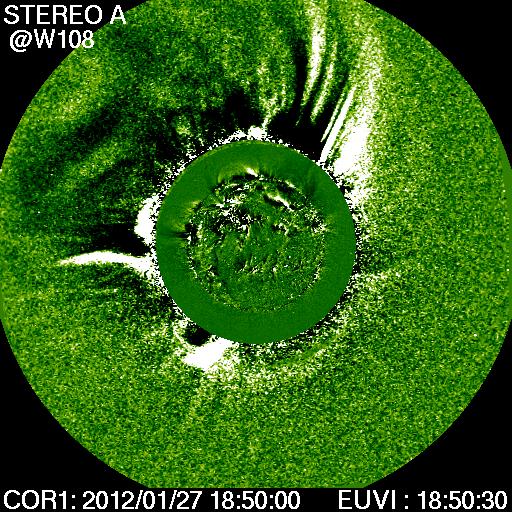}
   \includegraphics[width=0.45\textwidth]{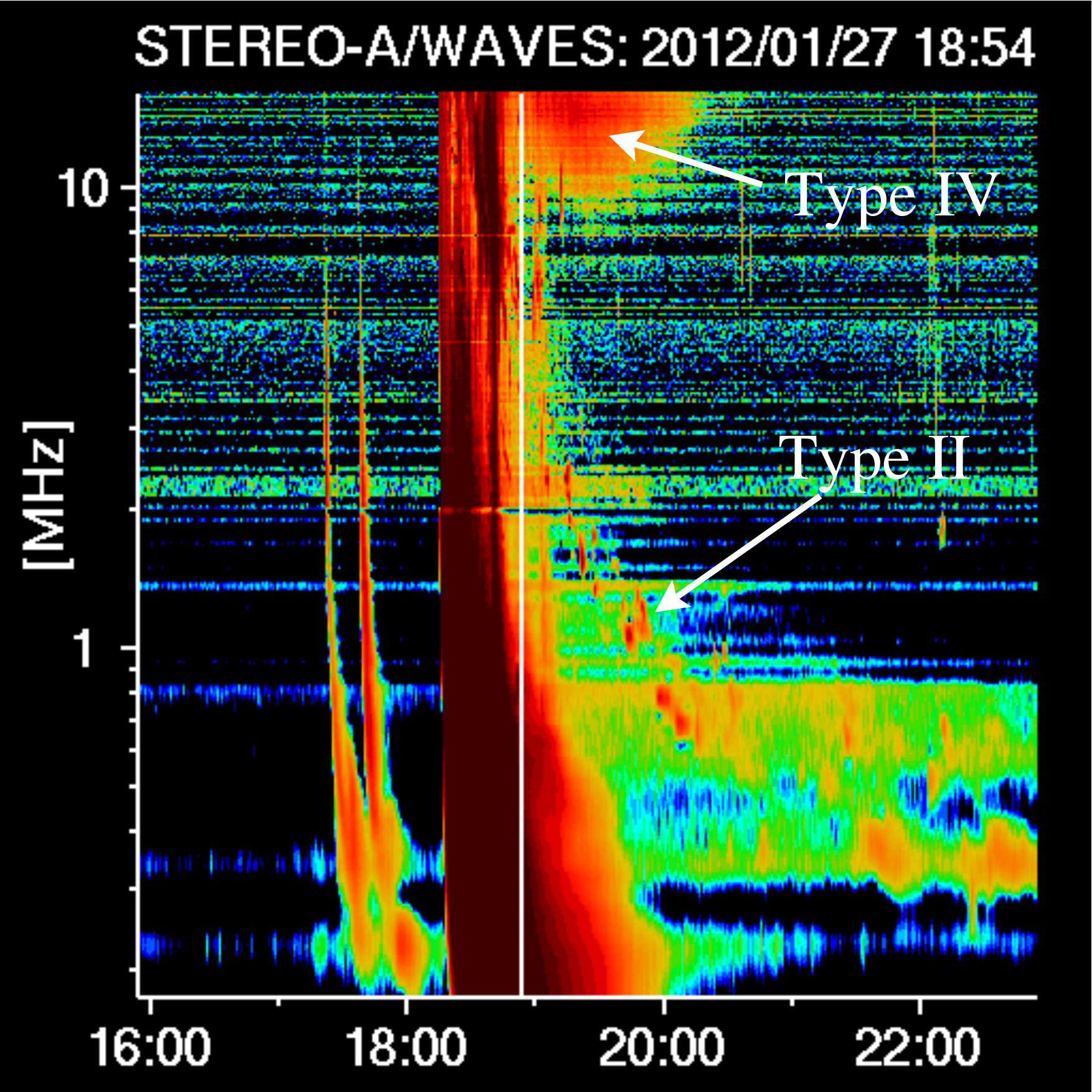}
 \includegraphics[width=0.45\textwidth]{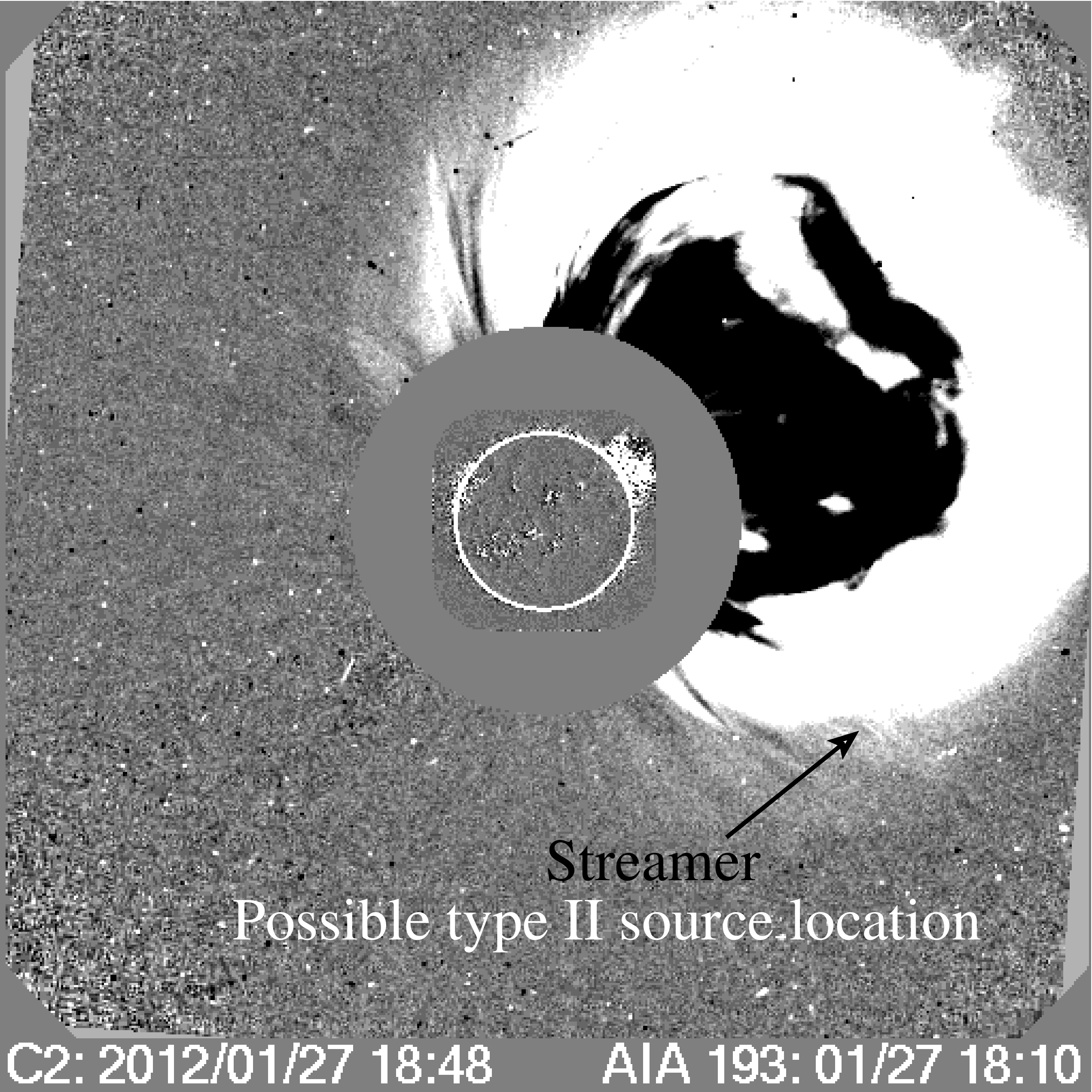}
 \includegraphics[width=0.45\textwidth]{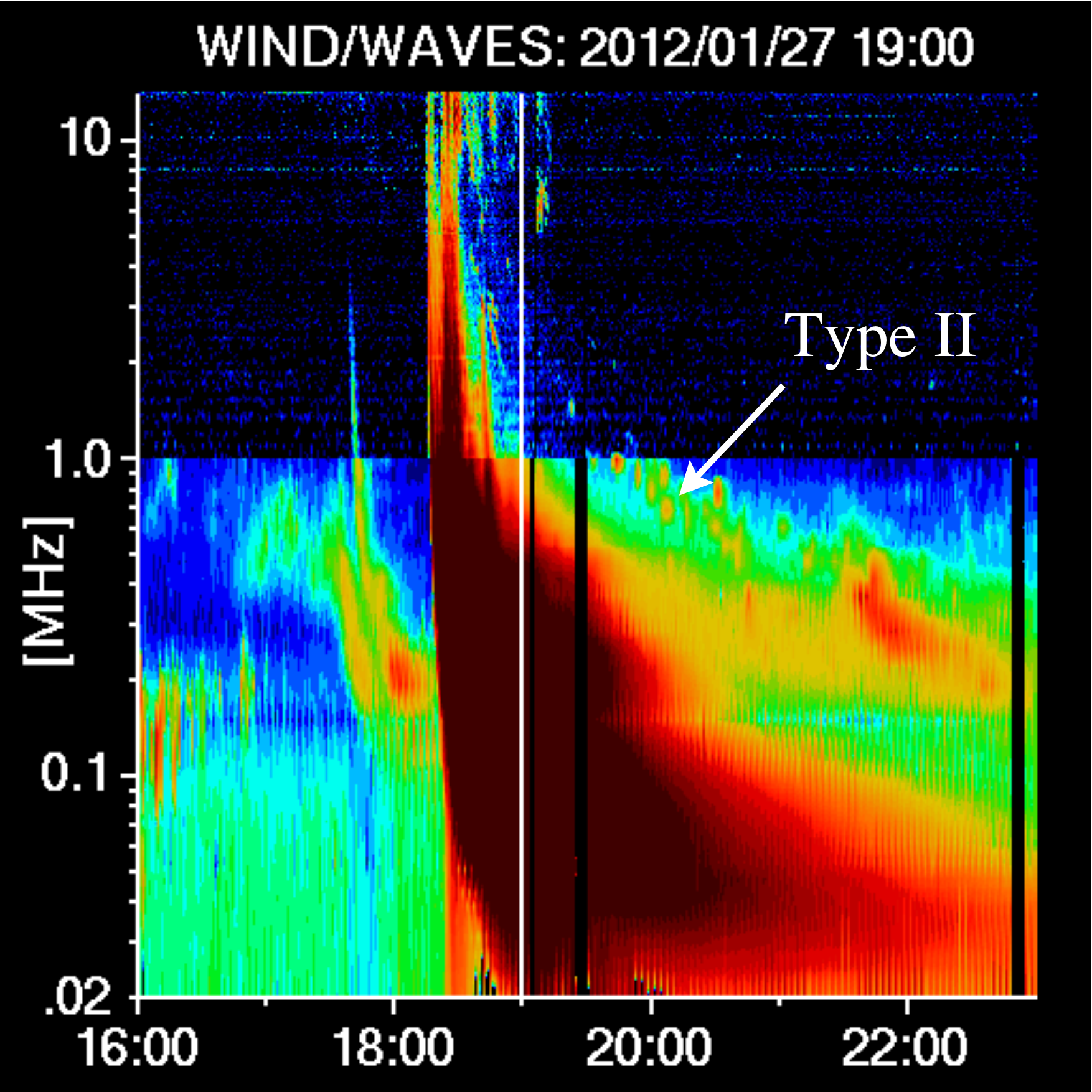}
  \includegraphics[width=0.45\textwidth]{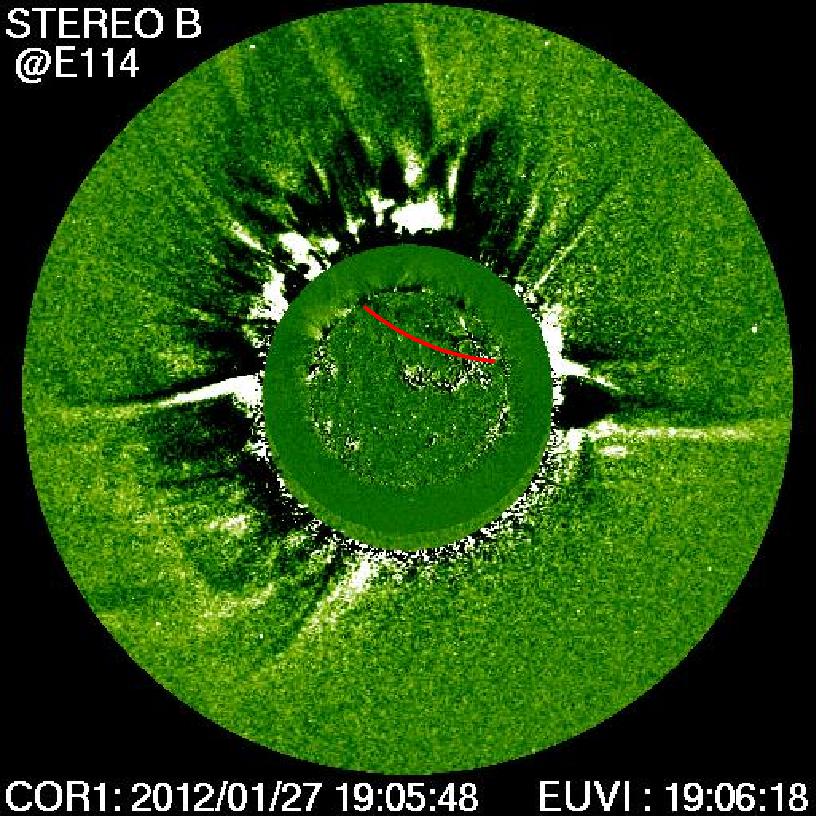}
  \includegraphics[width=0.45\textwidth]{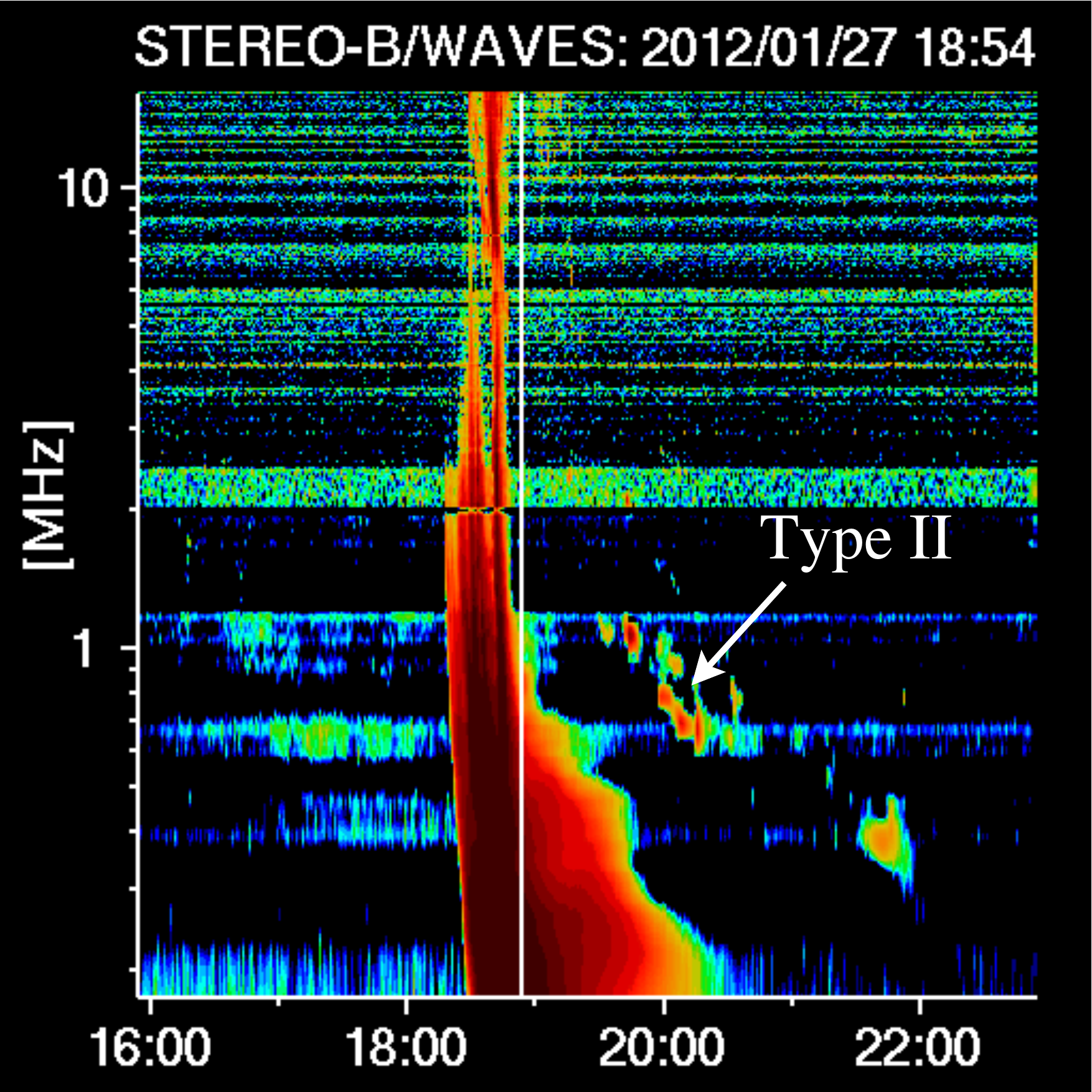}
  \caption{Solar event on 27 January 2012: Flare location at N27W71 in AR1402.
    The red line in the STEREO-B EUVI difference image indicates the maximum extent
    of the EUV wave on the visible disk.
} 
\label{2012jan27}%
\end{figure}
\begin{figure}[!ht]
   \centering
 \includegraphics[width=0.45\textwidth]{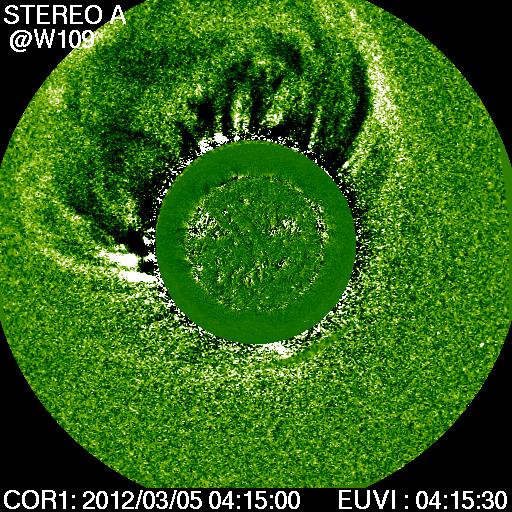}
  \includegraphics[width=0.45\textwidth]{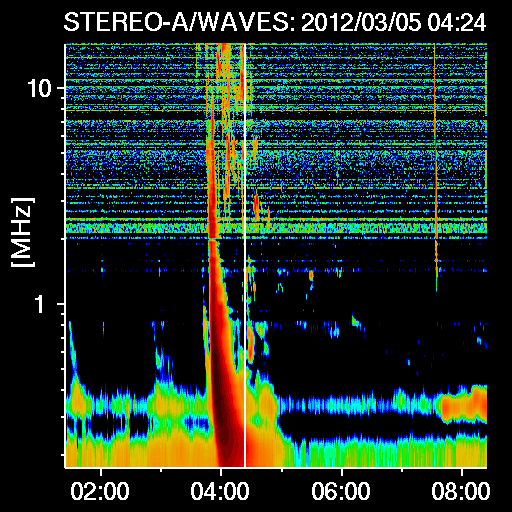}
  \includegraphics[width=0.45\textwidth]{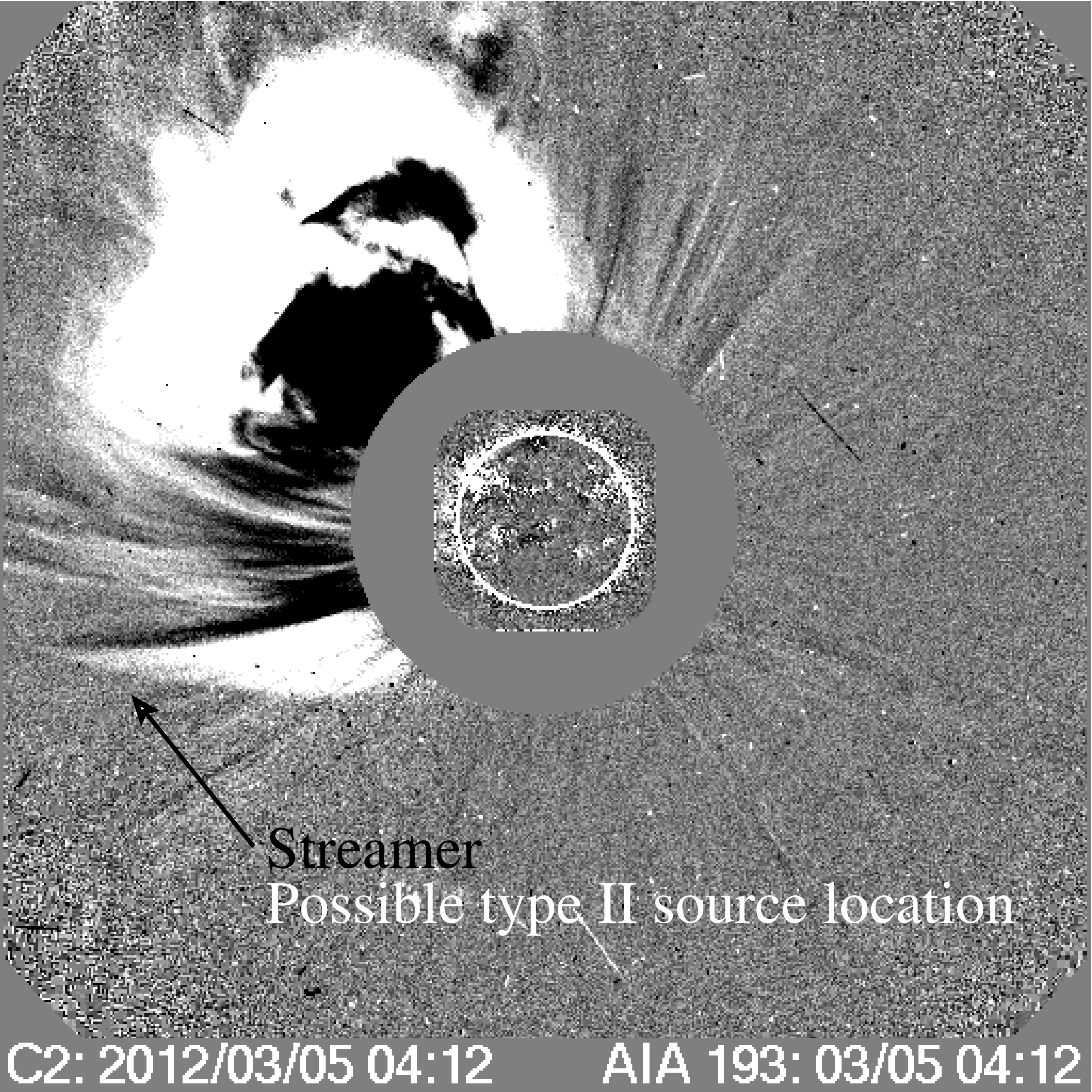}
 \includegraphics[width=0.45\textwidth]{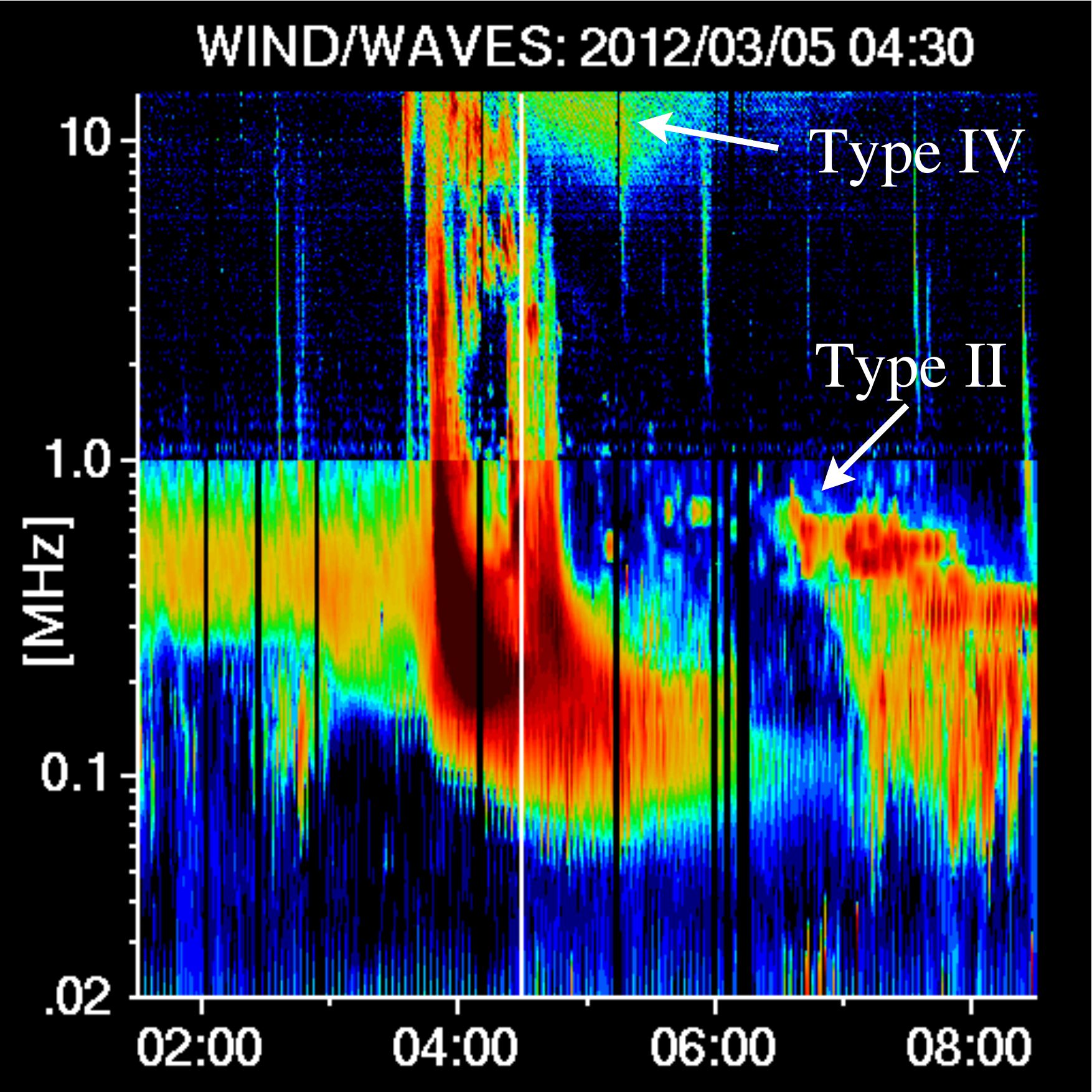}
  \includegraphics[width=0.45\textwidth]{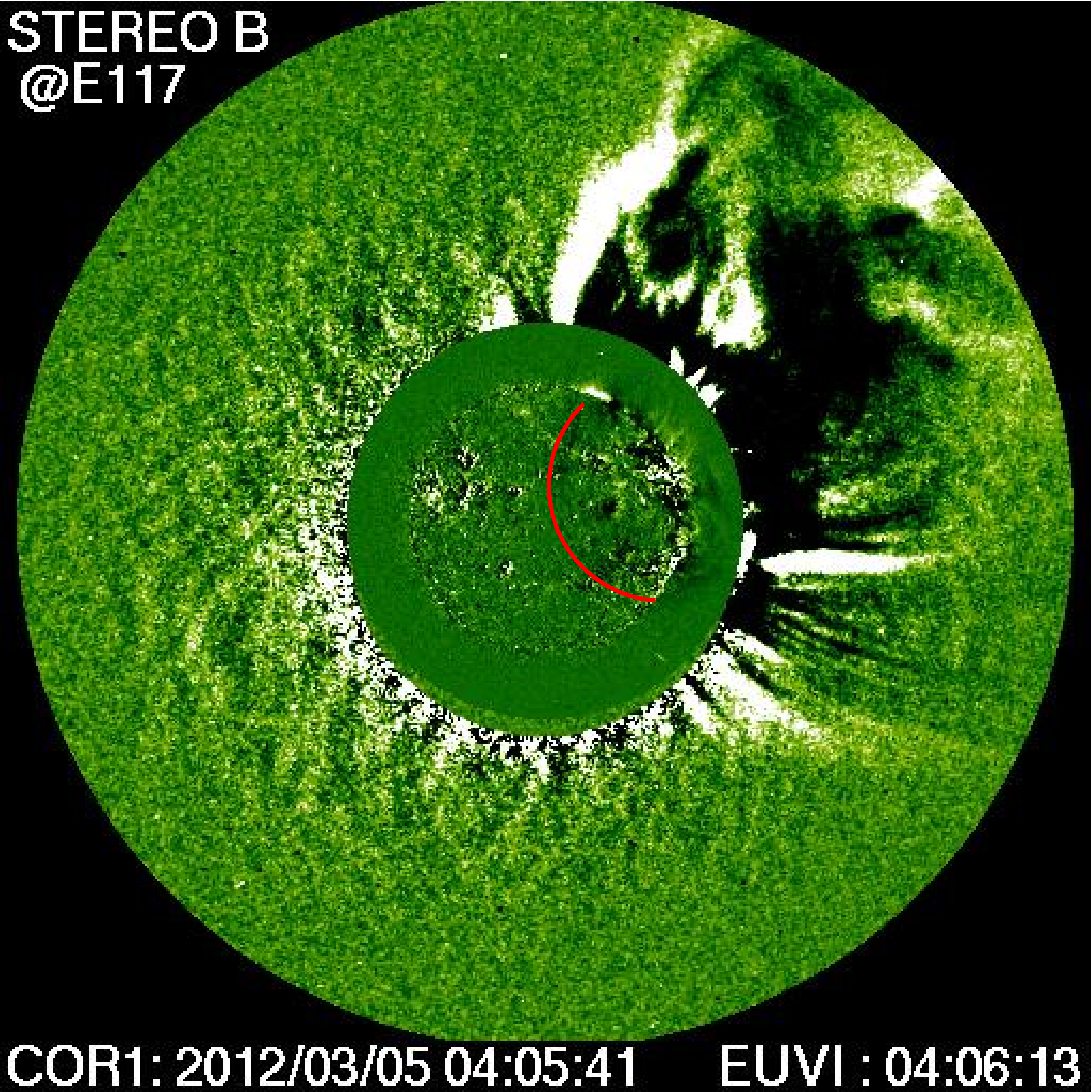}
   \includegraphics[width=0.45\textwidth]{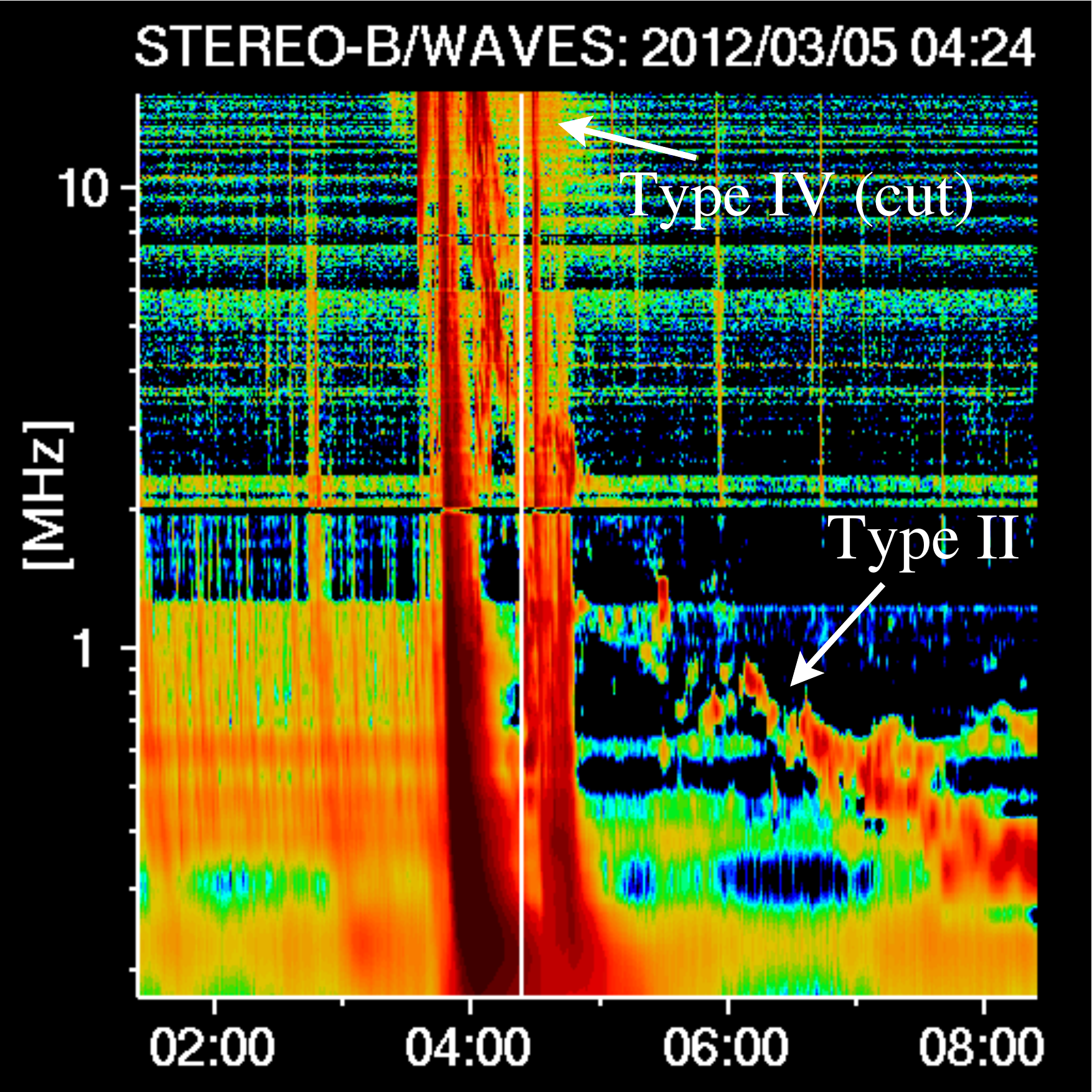}
   \caption{Solar event on 5 March 2012: Flare location at N17E52 in AR1429.
     The red line in the STEREO-B EUVI difference image indicates the maximum extent
     of the EUV wave on the visible disk.
} 
\label{2012mar5}%
\end{figure}

The magnetic field structure near the eruption location on each day is shown
in Figure \ref{pfss}. Figures \ref{2011june4A} -- \ref{2012mar5} present the
radio dynamic spectra and the corresponding coronagraph and EUV difference
images for the five events, with the three different views from the three spacecraft. 

\pagebreak
\clearpage
\newpage

\pagebreak
\clearpage
\newpage

\section{Results}

\subsection{First Event on 4 June 2011}

The flare started approximately at 06:20 UT from the backside of the Sun, 
near N20W140. The halo CME was first observed by SOHO/LASCO at 06:48 UT
at height 2.58 R$_{\odot}$, propagating to north-west with a linear
fit speed of 1407 km s$^{-1}$ (LASCO CME Catalog). An EUV wave occurred
on the backside, observed by STEREO-A/EUVI, and it originated from the
active region located at $\sim$W50 in this field of view. The first brightening
phase of the EUV wave started at 06:30~UT, after which a dimming phase was
observed at 06:50 UT. The EUV wave crossed most part of the disk within
the STEREO-A view, but the wave was not observed on the disk in the
other field-of-views.

The DH type IV event became visible in the dynamic spectrum at 07:12 UT
at 16 MHz, and it was observed by STEREO-A but not by {\it Wind} or
STEREO-B. At 08:00 UT it reached the lowest frequency of about 6~MHz.
As the flare site was at $\sim$W50 in STEREO-A view we cannot know if
a metric type IV burst existed, as radio emission originating from low
heights would have been blocked by the solar disk toward the Earth.

The type IV burst was atypical in a sense that it showed type III-like
burst structures superposed, and starting, from the wide-band emission
(Figure \ref{2011june4A}, top right dynamic spectrum). This indicates that
several open field lines existed along which particles could stream and/or escape
from the trap, see the potential magnetic field lines plotted in
Figure \ref{pfss}a.

A type II burst was observed by STEREO-A and {\it Wind}, but not by STEREO-B.
As the most probable location for the type II burst source is in the
western flank of the CME (Earth view, Figure \ref{2011june4A} middle left
difference image), it would have been behind the Sun in STEREO-B view and
hence undetectable. 

\subsection{Second Event on 4 June 2011}

The next type IV event occurred on the same day, 15 hours later, on 4 June 2011.
The flare originated from the same active region on the backside of the Sun,
starting at 21:45 UT and located at N20W160 approximately ($\sim$W70 in
STEREO-A view). The magnetic field structures could be as in the
earlier event on the same day (Figure \ref{pfss}a), but as the active region had
now rotated well behind the limb, we cannot be sure of the actual configuration.
The halo CME was first observed by SOHO/LASCO at 22:05~UT, at height 4.35 R$_{\odot}$,
with a linear fit speed of 2425 km s$^{-1}$.

An EUV wave that was followed by a dimming originated from the flare location
and it was well-observed by STEREO-A. The wave was visible near the limb in the
STEREO-B view, but no wave signatures were observed from the Earth view
(Figure \ref{2011june4B}, difference images on the left).

The DH type IV burst was observed to start at 22:15 UT, observed by
STEREO-A only, and it reached the lowest frequency of 6 MHz at 23:05~UT.
A very fast-drifting type II burst was observed by STEREO-A, but this
emission was not detected by the other instruments (Figure \ref{2011june4B},
top right spectrum). The type II burst source is again estimated to be located
in the western flank of the CME (Earth view), but this time in a more narrow
region that is blocked in view from the other radio instruments.


\begin{figure}[!h]
\centering
\includegraphics[width=0.8\textwidth]{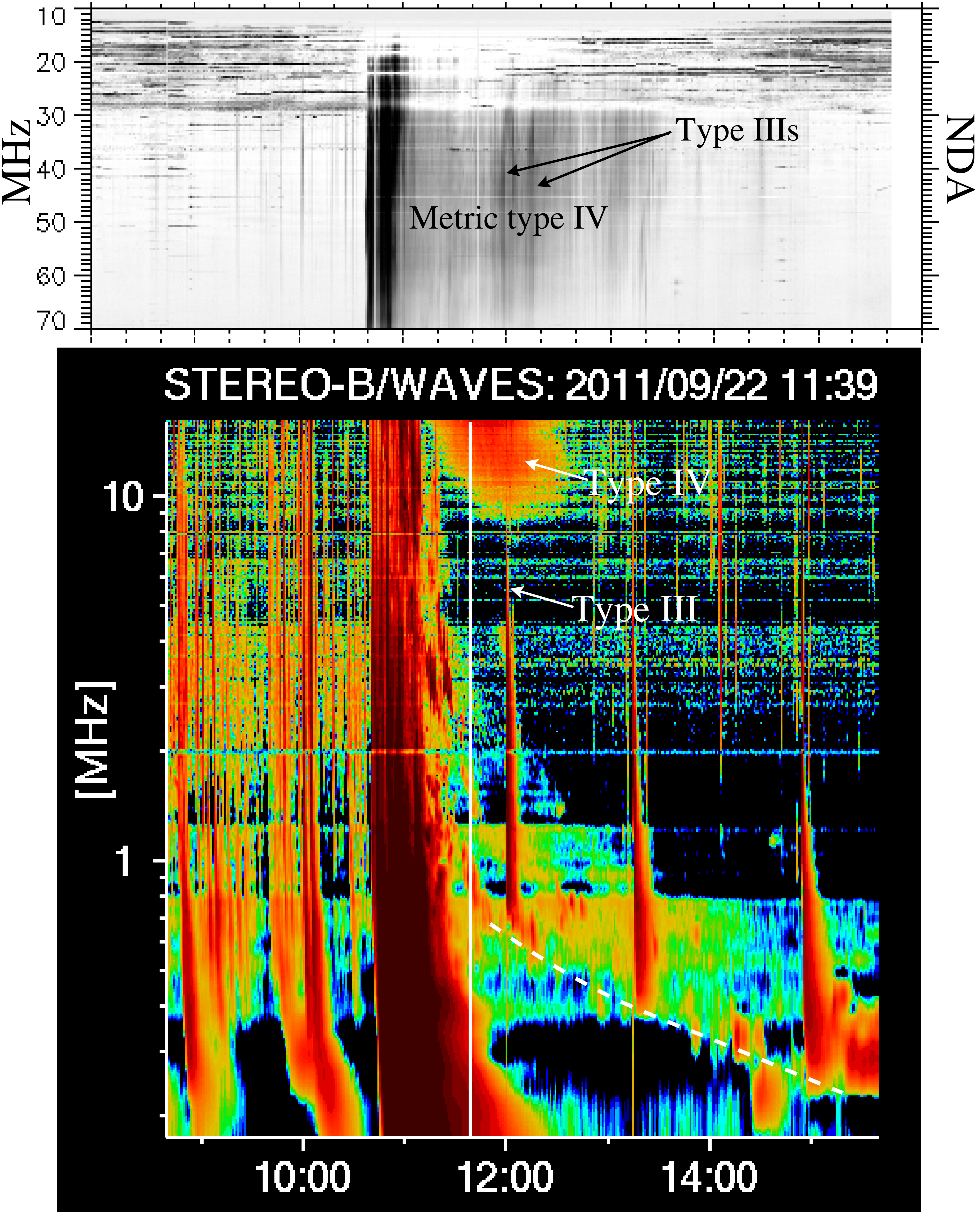}
\caption{Radio dynamic spectrum observed by STEREO-B/WAVES on
22 September 2011 (bottom) and {\it Nancay Decameter Array} (NDA) spectrum
at 70--10 MHz (top).
The flaring region was located at E80 in NDA view and at W10 in STEREO-B view.
The metric type IV emission observed by NDA was not detectable in the
{\it Wind}/WAVES data (same field-of-view, frequencies below 14 MHz).
The DH type IV burst observed in the STEREO-B field-of-view does not have
a one-to-one match in duration with the metric type IV burst.
The type III bursts stop at the type II burst emission lane (indicated by
a white dashed line), suggesting that the electron beams cannot pass the
type II shock front.} 
\label{nda20110922}
\end{figure}

\subsection{Event on 22 September 2011}

A GOES X1.4 class flare was observed to start at 10:29 UT on 22 September 2011
in active region NOAA 11302, located at N13E78. It reached maximum flux at 11:01 UT.
A halo CME was first observed by SOHO/LASCO at 10:48 UT, at height 2.98 R$_{\odot}$,
and it had a linear fit speed of 1905 km s$^{-1}$. An EUV wave related to this event
was mainly observed by STEREO-B/EUVI on the disk (Figure \ref{2011sep22}, bottom left
difference image), but it was also observed to propagate toward the west in Earth view,
with a speed of 595 km/s \citep{nitta2013}.

The DH type IV burst was observed at 11:10 UT by STEREO-B/WAVES at 16 MHz and it
reached the lowest frequency of 8 MHz at 12:15 UT. The DH type IV burst was preceded
by a decimetric-metric type IV burst observed by NDA (Earth view). These
bursts do not seem to be directly associated, as the metric type IV emission continued
after the DH emission had ended, see Figure \ref{nda20110922}. Also the drift
rates of the emission envelopes do not look to match. The metric type IV burst
shows groups of type III bursts with varying frequency drifts (both positive and
negative), similar to those reported by \cite{melnik2018}.

Near 12:00 UT all the three space  instruments recorded a type III burst in
their dynamic spectra. The type III burst (formed by a propagating electron beam)
is observed to cross the DH type IV burst in STEREO-B spectrum. This, and also
the later type III bursts, stop near the type II emission lane, suggesting that
the electron beams cannot pass the type II shock front (Figure \ref{nda20110922}).
The location of the type II burst source is most probably in the south-eastern
flank of the CME (Earth view, Figure \ref{2011sep22} middle left difference image),
making it observable to STEREO-B and {\it Wind}. Figure \ref{pfss}b shows some of
the open field lines along which electrons could stream out from the active region
and they are mostly directed toward STEREO-B.

\subsection{Event on 27 January 2012}

The active region NOAA 11402 produced an X1.7 GOES class flare at location N27W71 that
started at 17:37 UT and peaked at 18:37 UT on 27 January 2012. The flare was
associated with a halo CME that was first observed by SOHO/LASCO at 18:27 UT
at height 3.76 R$_{\odot}$, and had a linear fit speed of 2508 km s$^{-1}$. The
event was associated with an EUV wave that was observed globally on the disk
by STEREO-A/EUVI, and partly on the disk by STEREO-B/EUVI and SDO/AIA. The EUV
wave speed from Earth view was determined as 635~km~s$^{-1}$ \citep{nitta2013}.

The DH type IV burst was observed at 16 MHz at 18:30 UT by STEREO-A, but not by
the other instruments. The burst reached the lowest frequency of 9 MHz at
19:30 UT. The DH type IV burst was preceded by a metric type IV burst at
18:15--18:35 UT, observed by the {\it Green Bank Solar Radio Burst Spectrometer}
(GBSRBS) down to 300 MHz and by the {\it Radio Solar Telescope Network} (RSTN)
at 180--25 MHz. Due to the short duration of the metric type IV burst it is not
possible to determine if these two bursts were related. 

A type II burst was recorded by all the three space spectrographs, although the
type II emission lane was more patchy and narrow in the STEREO-B spectrum
(Figure \ref{2012jan27}). The most probable location for the type II burst source
is in the south-western flank of the CME (Earth view, Figure \ref{2012jan27} middle
left difference image), thus making it visible in all three field-of-views.
No type III bursts were observed during the DH type IV burst emission, although
some open field lines existed, directed to the north-west (Figure \ref{pfss}c).

\begin{figure}[!h]
\centering
\includegraphics[width=0.8\textwidth]{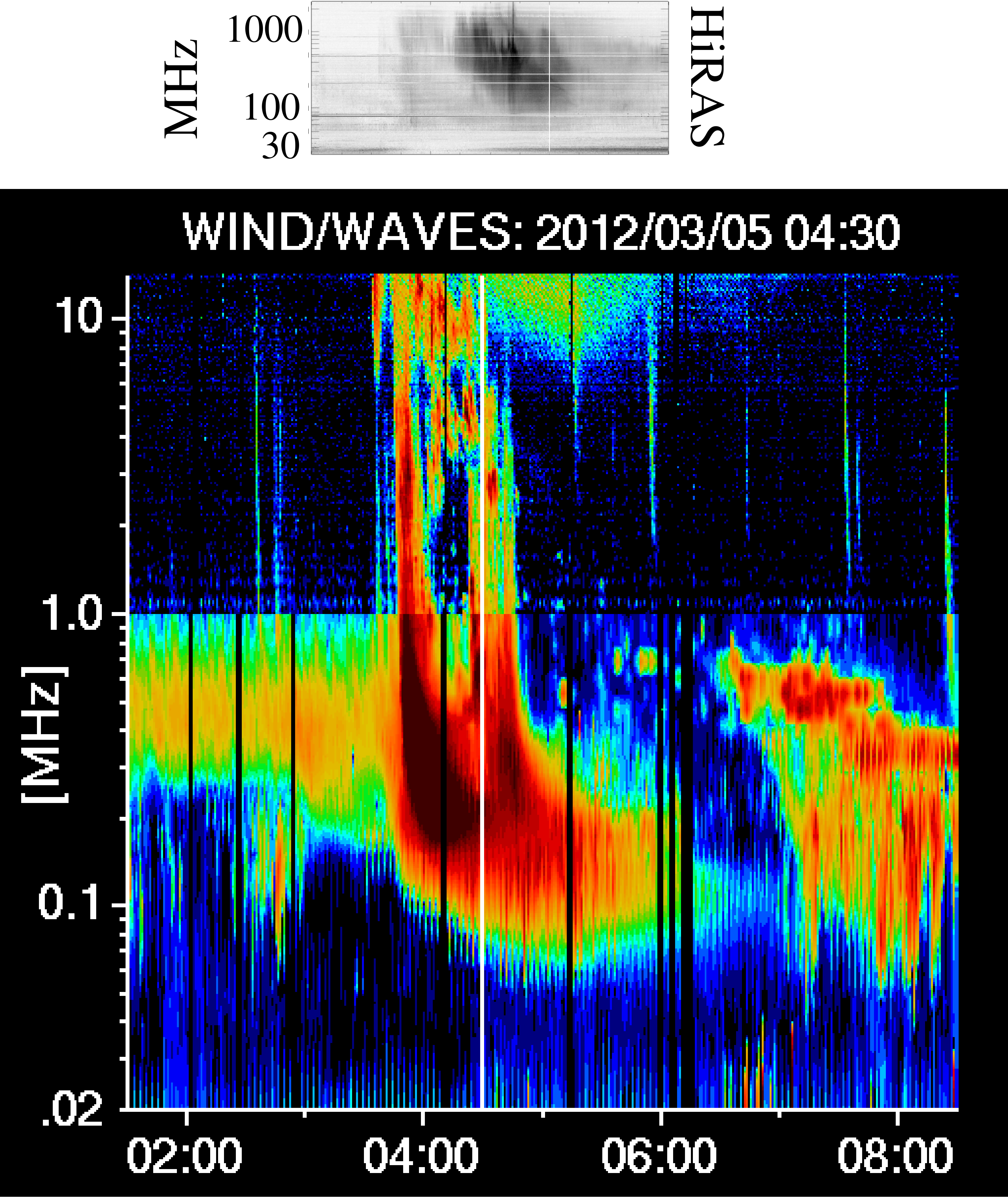}
\caption{Radio dynamic spectrum observed by Wind/WAVES on 
5 March 2012 (bottom) and by HiRAS at 2000--20 MHz (top).}
\label{hiras20120305}
\end{figure}

\subsection{Event on 5 March 2012}

An X1.1 GOES class flare started at 02:30 UT from the NOAA active region 11429
located at N17E52. At 03:12 UT a filament eruption formed into a CME, first visible
in the SOHO/LASCO images at height 3.0 R$_{\odot}$. The linear fit speed of
this CME was 594 km s$^{-1}$. A fast halo CME appeared at 04:00 UT at height
4.2 R$_{\odot}$, with a linear fit speed of 1531 km s$^{-1}$, overshadowing the
earlier CME. The launch of the second CME was most probably related to the fast
rise in X-ray flux, observed to start at 03:25 UT. An EUV wave was observed
globally on the disk from Earth view, and it was estimated to move toward the
south-west direction at a speed of 915 km s$^{-1}$  \citep{nitta2013}. The wave
was partly observed also by STEREO-B/EUVI (Figure \ref{2012mar5}, bottom left
difference image). 

The DH type IV burst was observed as a compact event by {\it Wind}/WAVES,
starting at 04:15 UT and reaching the lowest frequency of 7 MHz at 05:15~UT.
It was also observed as shorter-duration ('cut') emission in the
STEREO-B view (Figure \ref{2012mar5}, bottom right dynamic spectrum).
This sudden end of emission at all the type IV burst frequencies suggests that
the emission was blocked from being observed, rather than ending as an
emission process. No significant type III burst activity was observed during
the type IV burst emission. Some open field lines are however visible
in the PFSS maps, directed toward the north-east, see Figure \ref{pfss}d.  

At decimeter--meter waves drifting continuum emission was observed at
03:30--05:15 UT (Figure \ref{hiras20120305}, on top, spectrum from
NICT/HiRAS). This could be a faint type IV burst or a noise
storm. HiRAS spectrum shows that the emission envelope has a drift
toward the lower frequencies, with a drift and timing that quite well
matches those of the DH type IV burst.

The {\it Wind} and STEREO-B spectra show also intense type II burst
emission. This event has been analysed by \cite{magdalenic2014}, who
used radio triangulation technique to locate the type II burst source
positions. These were found to be close to the south-eastern flank of
the CME (Earth view, Figure \ref{2012mar5}, middle left difference image),
where streamers were also located. The appearance of the type II emission
was hence suggested to be due to shock wave and streamer interaction.


\begin{figure}[!b]
   \centering
   \includegraphics[width=0.8\textwidth]{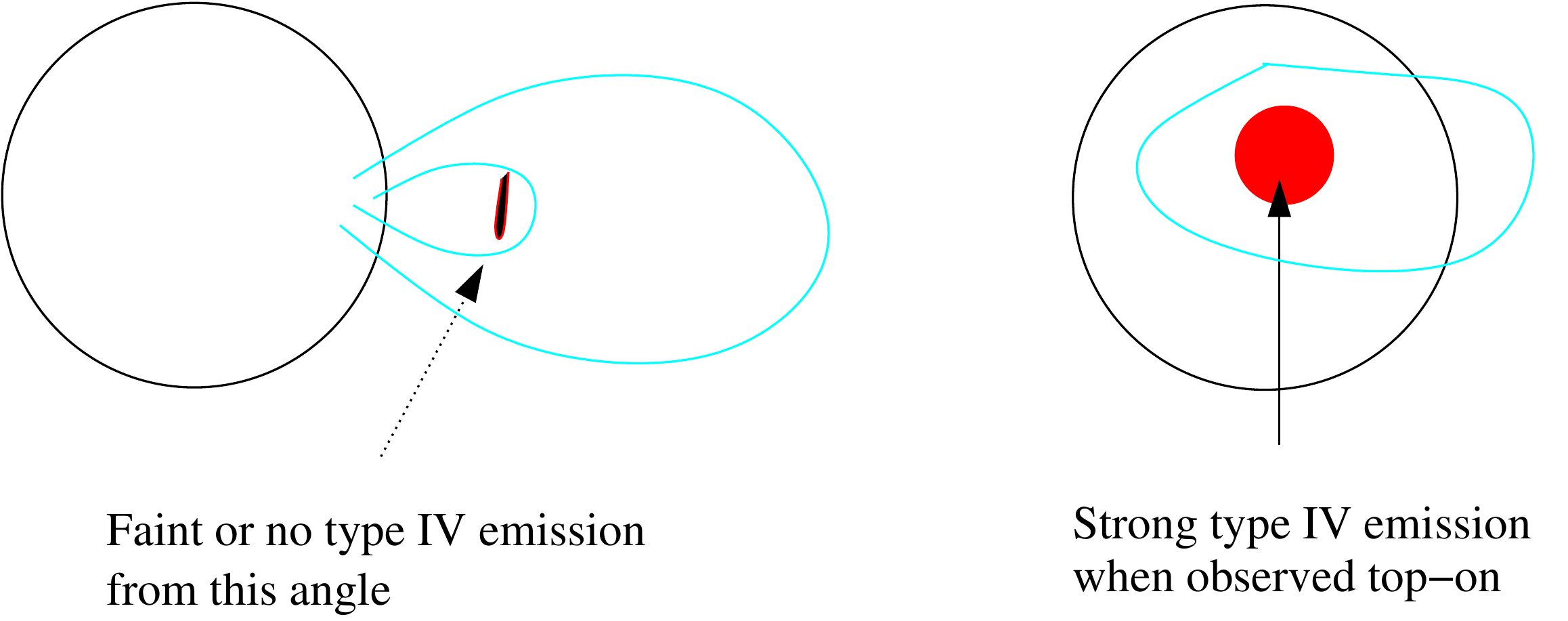}
   \caption{Schematic drawing along the lines presented in Kundu (1965)
     on where the type IV burst source may be located. Directivity could
     be related to the extent of the source, which is larger when viewed
     top-on, in the direction of propagation.} 
\label{type4source}%
\end{figure}

\section{Discussion} 

The first ideas to explain why type IV bursts show directivity in
their emission were reviewed by \cite{kundu65}, see especially Fig. 11-12
in the book and references therein. A schematic drawing in Figure \ref{type4source}
presents the suggested configuration, based on observations of metric
type IV bursts as observations from space at low frequencies did not yet exist.
A flare explosion ejects a column of gas and the outward moving gas carries
a magnetic field with it. A small fraction of particles accelerated to very
high energies can be trapped in the frozen-in magnetic fields and will
emit continuum (type IV) emission in a wide frequency range.
The directivity of emission could then be associated with the large extent
of the type IV source in the direction of motion. Later on, a segment of
the type IV burst source can separate and move outward, but it will no
longer emit synchrotron emission due to the insufficient magnetic fields.

\cite{gopal2016} recently studied a DH type IV burst on 7 November 2013 
that was observed to be intense and complete only in STEREO-B observations,
with STEREO-A seeing a partial burst and {\it Wind} no burst at all.
However, \cite{melnik2018} reported later that the type IV burst was
observed also at decameter waves by URAN-2, from Earth.
\cite{gopal2016} concluded that the type IV emission was directed
along a narrow cone, less than $\approx$\, 60\,degrees in width from
above the flare site. They suggested that the source of energy for the
burst was the flare: electrons accelerated due to flare reconnection
and then trapped in the post-eruption structures, producing radio
emission at the local plasma frequency.  As \cite{melnik2018} were
able to observe the start of the type IV burst at decameter waves,
they confirmed that the emission was indeed plasma emission since it
was highly polarized. \cite{melnik2018} suggested that the source
region for the type IV burst was the CME core but they did not specify
why URAN-2 could observe it while {\it Wind} could not.

\begin{figure}[!ht]
   \centering
   \includegraphics[width=0.8\textwidth]{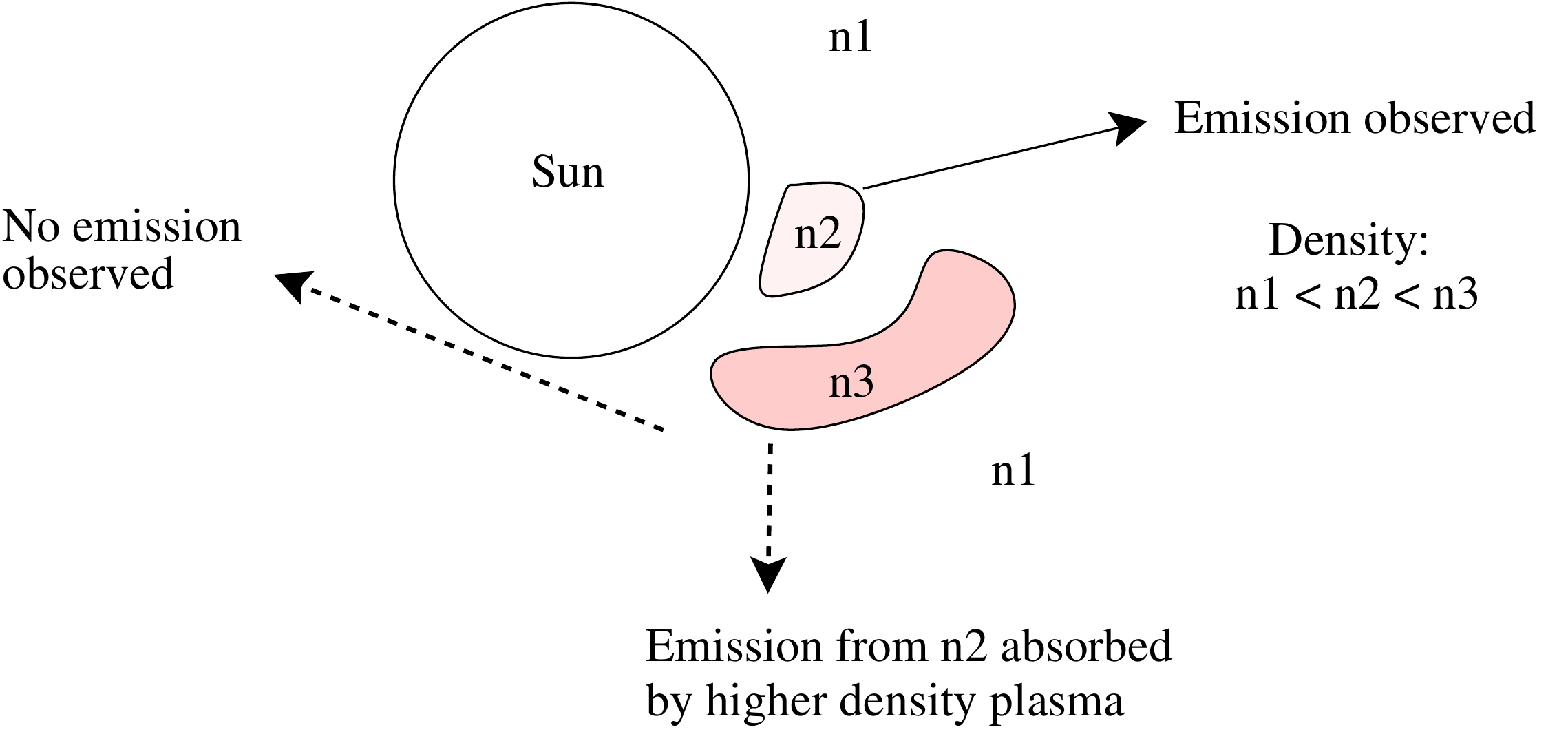}
   \caption{Schematic drawing for not detecting radio emission from a
       coronal source (n2) which is located behind a region of higher
       density plasma (n3). Due to the plasma frequency, longer wavelengths
       produced by plasma emission and/or sychrotron emission in the n2-region
       will be stopped and absorbed in the n3-region. To directions not blocked
       by the n3-region or the Sun itself, the emission can be observed in full.}
\label{plasmaemission}%
\end{figure}

We note that it is possible that the observed directivity may not be
caused by the emission mechanism and/or the extent of the type IV source.
Alternatively, the produced radiation could be stopped in certain directions,
by for example higher density plasma if it existed in between the source
and the observer. In the solar corona such local density enhancements can
occur in coronal streamers and shock-related compression fronts \citep{kwon2018}.
Figure \ref{plasmaemission} shows how emission from the n2-region cannot
pass through the higher density plasma in the n3-region, and hence it
cannot be observed along this direction. The dense solar atmosphere
can also block part of the emission, especially if the source region is
located on the backside of the Sun and the direction of propagation is
away from the observer.

The five events in our study presented configurations similar as in 
Figure \ref{plasmaemission}: a CME source region was located in n2-region  
and a streamer was located in n3-region. Moreover, in all events a type II burst
was estimated to be located near the streamer, as a CME flank shock. Therefore,
a higher density region existed toward a direction where the type IV burst
could not be observed.
In the two 4 June 2011 events the type II\,--\,streamer region was in the
south-western CME flank, blocking type IV emission toward the Earth but
being fully observable with STEREO-A. In the 22 September event the
type II\,--\,streamer region was in the south-eastern CME flank, and the type
IV burst was observed by STEREO-B but not by {\it Wind} from the Earth
direction. In the 27 January 2012 event the type II\,--\,streamer region was in
the south-western CME flank, with a similar configuration as in the 4 June 2011
events. In the 5 March 2012 event the type II\,--\,streamer region was located in the
south-eastern CME flank (projected Earth view), but the CME propagation
direction indicated that the type IV burst source was directed more
toward the Earth and the streamer was located behind it. The effect was that
the type IV burst was observed by {\it Wind} near Earth, but it was only partly
visible for STEREO-B. 

In the 22 September 2011 event we observed several type III bursts
that ended near the type II burst emission frequency. This type of 'cut-off'
has been reported earlier by \cite{alhamadani2017}. One possible
explanation for the disappearance of type III radio emission could be
a reduced level of beam-driven Langmuir waves, like in the case when
two electron beams cross and the radio emission gets depleted
\citep{briand2014}. However, in this scenario the level of radio
waves is recovered after the beam crossing, unlike in our event on
22 September 2011.  This suggests that also the type III bursts 
could not pass the type II\,--\,streamer region, but were stopped.

\section{Summary and Conclusions}

In this paper, we have studied five solar events that showed intense
and compact DH type IV bursts, with the duration of around one hour for most
of them, but visible only from one viewing angle. The associated flares had high
intensities (three GOES X-class, two unclassified as they appeared on the
backside of the Sun). All the CMEs had very high speed and were halo-type.
Our analysis showed that EUV waves were observed in all of the events, and
intense and compact type IV bursts were observed only when the EUV wave
propagated globally across the whole visible disk. This may simply indicate
the propagation direction of the depleted CME material, if not the directivity
of the type IV radio emission.

All the CMEs were associated with type II radio emission at DH wavelengths,
which indicates propagating shock fronts driven by CMEs. The calculated
estimates for the type II heights showed that the shocks were not bow
shocks at the leading fronts of the CMEs, but were located near the CME
flanks. In one event the type II location at the CME flank had also
earlier been verified with the radio triangulation technique. In all five
events there were high-density streamers present at the deduced locations
of the type II shocks. 
Shock--streamer interaction has been found to be the source of type II burst
emission in several earlier studies ({\it e.g.}, \citeauthor{alhamadani2017},
\citeyear{alhamadani2017}, and references therein). Our analysis showed
that in all five events the type IV emission was not observed from a direction
where the type II\,--\,streamer region was located in between the CME source
region and the observer.

We therefore conclude that a high density region existed in all events
toward the direction where the DH type IV burst could not be observed.
As the type II burst sources were found to be located higher in the corona than
the type IV sources, the type IV emissions could have been stopped in these 
shock regions. Dense type~II emitting regions could simply be optically
thick for the type IV emission. In one of the events the type II region was
also capable of stopping later-accelerated electron beams, visible as
type III bursts that ended near the type II burst lane.

\begin{figure}[!h]
\centering
\includegraphics[width=0.7\textwidth]{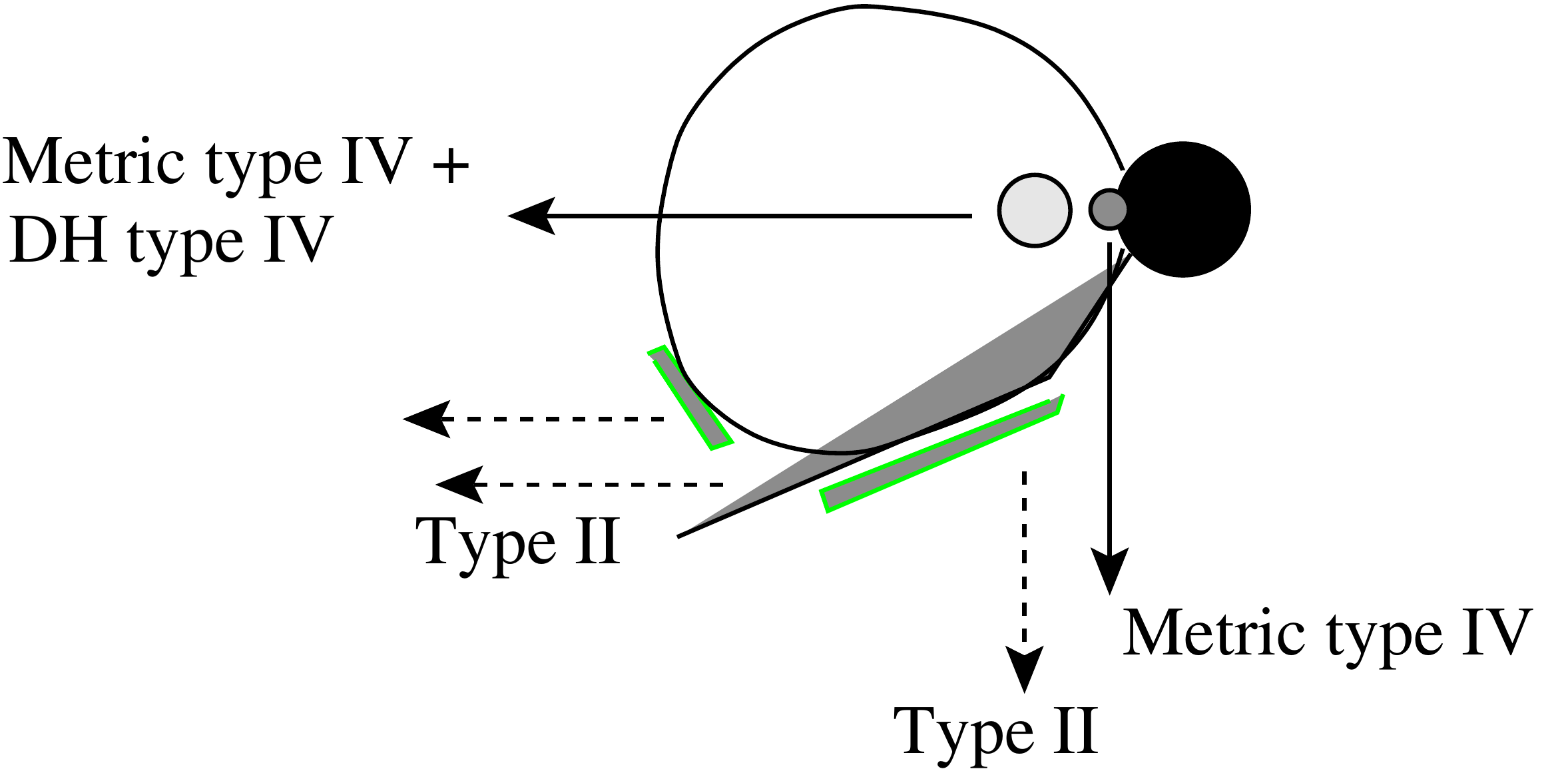}
\caption{Schematic cartoon that explains why metric type IV emission
  sources can be observed when DH type IV sources can not, from the same
  viewing angle. The green regions indicate possible type II burst
  locations and the filled grey region shows streamer position. In
  this scenario DH type IV emission is observed only from one viewing
  angle, because the plasma near the type~II emitting region is
  optically thick for the type IV emission to pass through. The higher
  density region coincides here with a type II shock\,--\,streamer region
  and it blocks the type IV emission only at longer DH wavelenghts,
    so that the metric type IV emission can still be observed.}
\label{type4model}
\end{figure}

To summarize our results and also to tie up our suggested scenario with
an earlier analysed event -- where a metric type IV burst could be observed
from a direction where the DH type IV burst could not -- we present a schematic
cartoon in Figure \ref{type4model}. Basically, a type IV burst can be observed
from a direction where nothing blocks or absorbs the radio waves. Based on
our analysis, possible blocking regions could be located at CME flanks, where
shock-streamer interactions may form type II radio bursts and where electron
densities can be much higher than in the surrounding space. These regions
may form only higher in the corona, making it possible for the metric
type IV bursts to be visible at lower coronal heights. We also note that in
most cases the type II bursts themselves are visible even if the type IV bursts
are not.


\begin{acks}
We thank the anonymous referee for comments and suggestions that helped to
improve this article.   
We are grateful to all the individuals who have contributed in creating 
and updating the various solar event catalogues. 
The CME catalog is generated and maintained at the CDAW Data Center by 
NASA and the Catholic University of America in cooperation with the 
Naval Research Laboratory. 
The {\it Wind} WAVES radio type II burst catalog has been prepared by 
Michael L. Kaiser and is maintained at the Goddard Space Flight Center.  
SOHO is a project of international cooperation between ESA and NASA. 
N. Talebpour Sheshvan wishes to thank CIMO (The Centre for International
Mobility, Finland) for financial support, contract TM-16-10298. 
\end{acks}

\vspace{5mm}

{\bf Disclosure of Potential Conflicts of Interest} 
The authors declare that they have no conflicts of interest.


\end{article}
\end{document}